\newif\ifconfver
\confverfalse      

\ifconfver
     \documentclass[10pt,twocolumn,twoside]{IEEEtran}
\else
    \documentclass[11pt,draftcls,onecolumn]{IEEEtran}
\fi

\usepackage{graphicx,booktabs,color}  
\usepackage{amsmath,epsfig,amsfonts,amssymb,graphics,psfrag,amsthm,calc,subfigure,url,bm,cite}
\usepackage{algorithm,algpseudocode}
\usepackage{float}
\usepackage{stfloats}  
\usepackage{cases}

\newlength{\twidth}
\ifconfver
\else
\fi

    \makeatletter
    \def\multilimits@{\bgroup
  \Let@
  \restore@math@cr
  \default@tag
 \baselineskip\fontdimen10 \scriptfont\tw@
 \advance\baselineskip\fontdimen12 \scriptfont\tw@
 \lineskip\thr@@\fontdimen8 \scriptfont\thr@@
 \lineskiplimit\lineskip
 \vbox\bgroup\ialign\bgroup\hfil$\m@th\scriptstyle{##}$\hfil\crcr}
    \def\Sb{_\multilimits@}
    \def\endSb{\crcr\egroup\egroup\egroup}
\makeatother

\definecolor{orange}{RGB}{255,107,0}

\newtheorem{Lemma}{Lemma}
\newtheorem{Prop}{Proposition}
\newtheorem{Theorem}{Theorem}
\newtheorem{Corollary}{Corollary}

\theoremstyle{remark}
\newtheorem{Remark}{Remark}

{\begin{list}{}%
    {\setlength{\rightmargin}{0pc}%
    \setlength{\leftmargin}{1pc} }} %
{\end{list}}

{\begin{list}{}%
    {\setlength{\rightmargin}{1pc}%
    \setlength{\labelsep}{0pc}
    \setlength{\leftmargin}{1pc} }} %
{\end{list}}

{\begin{list}{}%
    {\setlength{\rightmargin}{0pc}%
    \setlength{\leftmargin}{0pc} }} %
{\end{list}}

{\begin{list}{}{
    \settowidth{\labelwidth}{\mbox{\textnormal{#1}}}%
    \setlength{\leftmargin}{\labelwidth+\labelsep}}}%
{\end{list}}

\begin{document}

\bibliographystyle{IEEEtran}

\title{Spatially Selective Artificial-Noise Aided Transmit Optimization for MISO Multi-Eves Secrecy Rate Maximization
}

\ifconfver \else {\linespread{1.1} \rm \fi

\author{
Qiang Li$^\dag$ and Wing-Kin Ma$^\ddag$
\thanks{$^\S$ Copyright (c) 2012 IEEE. Personal use of this material is permitted. However, permission to use this material for any other purposes must be obtained from the IEEE by sending a request to pubs-permissions@ieee.org. This work is supported
by a Direct Grant awarded by the Chinese University of Hong Kong (Project Code 2050489).}
\thanks{$^\dag$Qiang Li is with
Department of Electronic Engineering, The Chinese University of Hong
Kong, Shatin, Hong Kong S.A.R., China. E-mail: qli@ee.cuhk.edu.hk.}
\thanks{$^\ddag$Wing-Kin Ma is the corresponding author. Address:
Department of Electronic Engineering, The Chinese University of Hong
Kong, Shatin, Hong Kong S.A.R., China. E-mail: wkma@ieee.org.}
}


\maketitle

\ifconfver \else
\begin{center} \vspace*{-2\baselineskip}
\end{center}
\fi
\begin{abstract}
Consider an MISO channel overheard by multiple eavesdroppers.
Our goal is to design an artificial noise (AN)-aided transmit strategy, such that the achievable secrecy rate is maximized subject to the sum power constraint.
AN-aided secure transmission has recently been found to be a promising approach for blocking eavesdropping attempts.
In many existing studies, the confidential information transmit covariance and the AN covariance are not simultaneously optimized.
In particular, for design convenience, it is common to prefix the AN covariance as a specific kind of spatially isotropic covariance.
This paper considers joint optimization of the transmit and AN covariances for secrecy rate maximization (SRM),
with a design flexibility that the AN can take any spatial pattern.
Hence, the proposed design has potential in jamming the eavesdroppers more effectively, based upon the channel state information (CSI).
We derive an optimization approach to the SRM problem through both analysis and convex conic optimization machinery.
We show that the SRM problem can be recast as a single-variable optimization problem,
and that resultant problem can be efficiently handled by solving a sequence of semidefinite programs.
Our framework deals with a general setup of multiple multi-antenna eavesdroppers, and can cater for additional constraints arising from specific application scenarios, such as interference temperature constraints in interference networks.
We also generalize the framework to an imperfect CSI case where a worst-case robust SRM formulation is considered.
A suboptimal but safe solution to the outage-constrained robust SRM design is also investigated.
Simulation results show that the proposed AN-aided SRM design yields significant secrecy rate gains over an optimal no-AN design and the isotropic AN design, especially when there are more eavesdroppers.
\\\\
\ifconfver
\else
\noindent {\bfseries EDICS}: MSP-CODR (MIMO precoder/decoder design), MSP-APPL (Applications of MIMO communications and signal processing), SAM-BEAM (Applications of sensor and array multichannel processing)
\fi
\end{abstract}

\ifconfver \else \IEEEpeerreviewmaketitle} \fi

\ifconfver \else
\newpage
\fi

\section{Introduction}

In the last decade, multi-antenna techniques have been extensively
investigated from the perspective of providing high throughput and reliable communications.
Recently, there has been growing interest in using multiple antennas to
achieve secure communication, which is known as {\it physical-layer
secrecy}. Intuitively speaking, the
idea of physical-layer secrecy is to add structured redundancy
in the transmit signal such that the legitimate user can correctly
decode the confidential information, but the eavesdroppers
can retrieve almost nothing from their observations~\cite{Wyner1975,Liang_book}. To
make physical-layer secrecy viable, we usually need the
legitimate user's channel condition to be better than the eavesdroppers'.
However, this may not be always possible in practice. To alleviate the dependence on the channels conditions, recent studies
are mainly focused on multi-antenna transmission, since
multiple transmit antennas provide spatial degrees of
freedom (d.o.f.) to degrade the reception of the eavesdroppers. A possible way
to do this is transmit beamforming, which concentrates the transmit signal over
the direction of the legitimate user while reducing power leakage to the eavesdroppers at the same time. Apart from this, a more active approach is to send artificially generated noise to interfere the eavesdroppers deliberately.

\begin{figure}[htp]
\centerline{\resizebox{.85\textwidth}{!}{\includegraphics{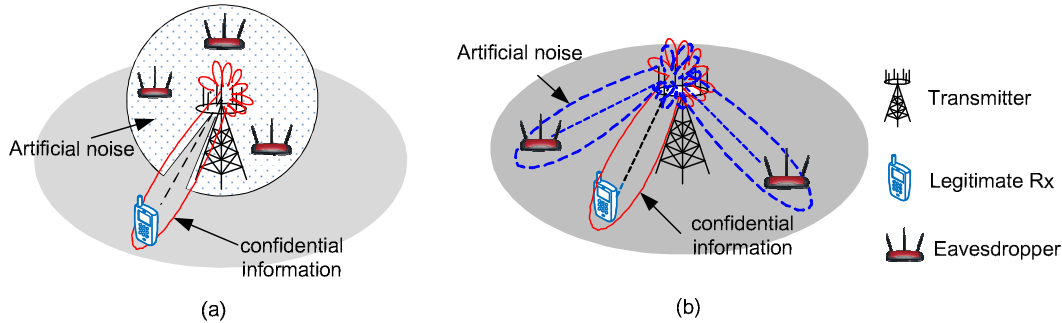}}
}
\vspace{-.5\baselineskip}
\caption{Secure transmission by (a) isotropic AN; (b) spatially selective AN.}
\label{fig:fig1}
\end{figure}

The notion of using artificial noise (AN) to enhance physical-layer security was first introduced by Negi and Goel
in~\cite{Negi2005}, and has received much attention in recent studies; see \cite{KW2007,Negi2005,Swindlehurst2009,Mukherjee2009,XYZHOU,Jorswieck,Jorswieck10} and the references therein.
The way of generating AN depends on how much the transmitter knows the eavesdroppers' channel state information (CSI).
Consider a case where no eavesdropper's CSI is available.
A popular design in such a case is {\it isotropic AN}~\cite{Negi2005},
where AN is uniformly spread on the legitimate channel's nullspace.
By doing so, one can guarantee that no interference will be made to the legitimate receiver, while the eavesdroppers' reception may be degraded by AN.
A picture is shown in Fig.~\ref{fig:fig1}(a) to illustrate how the isotropic AN design works.
On the other hand, consider cases where the eavesdroppers' CSI is available.
This may arise from scenarios where the eavesdroppers are also users of the system, and the transmitter aims to provide different types of users with different services.
Moreover, for an active eavesdropper,
the CSI can be estimated from the eavesdropper's transmission.
More interestingly,
a very recent study has suggested that even for a passive eavesdropper,
there is a possibility for one to estimate the CSI through the local oscillator power inadvertently leaked from the eavesdropper's receiver RF frontend~\cite{Mukherjee12}.
With knowledge of CSI, we can block the eavesdroppers much more effectively by generating spatially selective AN, rather than keeping AN isotropic~\cite{Ali2011,Liao10}.
Fig.~\ref{fig:fig1}(b) shows a picture to illustrate the idea of spatially selective AN.
However, perfect CSI may not be always available in practice,
and an important issue is how to robustify a secure transmit design in the presence of imperfect CSI, which is a more general and realistic assumption.
Tackling imperfect CSI in physical-layer security is presently an emerging subject with several concurrent endeavors; e.g.,
the worst-case robust design \cite{JHuang12,Wolf10,QLI2011}, the outage robust design \cite{Gerbracht12,JLI2012} and the ergodic design \cite{Jorswieck}.


\vspace{-.5\baselineskip}

This paper concentrates on the problem of AN-aided secrecy rate maximization (SRM) of a multi-input single-output (MISO) channel overheard by multiple multi-antenna eavesdroppers, with either perfect or imperfect CSI.
This problem has been addressed when there is no AN~\cite{QLI2011} (also \cite{KW2007} for the one eavesdropper case).
In the present problem, we consider joint optimization of the confidential information covariance and the AN covariance,
and there is a design flexibility for the AN to take any spatial pattern.
This AN-aided SRM problem is a challenging optimization problem.
The main difficulty lies in the AN covariance, which
makes the secrecy rate expression more complicated to optimize.
To make the AN-aided SRM problem easier to handle, a vast majority of existing works have to impose additional restrictions on AN, which leads to tractable but SRM suboptimal designs.
For example,
\cite{KW2007,Negi2005,Swindlehurst2009,Mukherjee2009,XYZHOU,Jorswieck,Gerbracht12,JHuang12,JLI2012} restrict AN in the nullspace of the legitimate channel,
\cite{Ali2011} requires AN to cause no decrease in the legitimate channel's mutual information,
and \cite{Jorswieck10,Jli2011,GanZheng11} assume that transmit beamforming is employed to generate AN.
We should point out that all the above mentioned works consider only one eavesdropper and the sum power constraint.
On the other hand,
there are works that consider additional design constraints for satisfying some application-specific requirements,
e.g., per-antenna power constraints, and the interference temperature constraints for interference networks~\cite{yiyangpei2010,Pei2011,Gursoy2010}.
However, AN is not incorporated in those designs and only one single-antenna eavesdropper is assumed.
In addition, there are some other AN-aided secure transmit designs, e.g., the SINR-based design~\cite{Swindlehurst2009,Liao10} and the MSE-based design~\cite{MinyanPei12}.
However, the goal of those designs is to provide the legitimate receiver with certain QoS, rather than directly maximizing the secrecy rate.

In this paper, we will develop a semidefinite program (SDP)-based optimization approach to handle the AN-aided SRM problem, with no structural restrictions on the AN.
Our problem formulation considers multiple multi-antenna evesdroppers,
either perfect or imperfect CSI with the eavesdroppers,
and additional design constraints arising from certain application-specific scenarios (such as the aforementioned).
Our main contributions are summarized as follows.

\begin{enumerate}

\item For the perfect CSI case, we derive an equivalent problem of the SRM problem through analysis.
The equivalent problem is less complex than the original SRM.
We show that the equivalent problem can be recast as a single-variable optimization problem, and that the latter can be handled by solving a sequence of convex SDPs,
for which efficient and reliable solvers are readily available~\cite{Sturm1999,Grant2011}.

\item For the imperfect CSI case, we consider a worst-case robust extension of the SRM problem. We show that the worst-case robust SRM (WCR-SRM) problem can be handled in a similar manner as SRM,
though
the development is more involved and a specific matrix inequality lemma is required.
A suboptimal but safe solution to an outage-constrained robust SRM (OCR-SRM) problem is also investigated.

\item Our SRM problem formulation assumes general transmit covariance for the confidential information,
and does not fix the transmit strategy as transmit beamforming.
Interestingly, in deriving the equivalent SRM problem,
we show that {\it transmit beamforming is an SRM optimal strategy for the confidential information transmission.}
This result is meaningful in giving a theoretical justification for using transmit beamforming in the considered scenario.
Moreover, the result applies to both the perfect CSI case and worst-case robust imperfect CSI case.
We should mention that the optimality of transmit beamforming has been proven in \cite{Shafiee2007,KW2007} under the assumption of
one eavesdropper, no AN and perfect CSI.
Our result in comparison is more general.

\end{enumerate}


This paper is organized as follows. A system model description
and problem statement is given in Section II. Section III
considers the SRM problem for the MISO multi-eavesdropper
scenario with perfect CSI, wherein the SDP-based optimization approach is established. Section IV extends the SRM results to the imperfect
CSI case. Simulation results comparing the proposed SRM
solutions and some other suboptimal secrecy transmit designs
are illustrated in Section V. Section VI concludes the paper.

Our notations are as follows. $\mathbf{A}^H$,
$\text{Tr}(\mathbf{A})$, rank(${\bf A}$) and $\det(\mathbf{A})$ represent the
Hermitian (conjugate) transpose,
trace, rank and
determinant of a matrix $\mathbf{A}$;
$\mathbf{I}$ denotes an identity matrix; $\| \cdot \|$ and $\|\cdot\|_F$ represent the
$\ell_2$ norm and Frobenius norm, respectively; $\mathbf{A}\succeq \mathbf{0}$ $({\bf A}\succ {\bf
0})$ means that $\mathbf{A}$ is a Hermitian positive semidefinite
(definite) matrix; $\left[{\bf A} \right]_{m,n}$ denotes the $(m,n)$th element of matrix ${\bf A}$; $\mathbb{R}_{+}$ denotes the set of all nonnegative
real numbers;
$\mathbb{H}_{+}^{N}$ denotes the set of all $N$-by-$N$ Hermitian positive semidefinite matrices;
${\bf x} \sim \mathcal{CN} ({\bm \mu}, {\bm
\Omega})$ means that ${\bf x}$ is a random vector following a
complex circular Gaussian distribution with mean ${\bm \mu}$ and
covariance ${\bm \Omega}$.

\section{System Model and Problem Statement}
\label{sec:prob_set}


\subsection{System Model}
\label{sec:sys_model}
Consider the scenario shown in Fig.~\ref{fig:fig1}(b).
A multi-antenna transmitter, called {\it Alice}, intends to send confidential information to a single-antenna legitimate
receiver, called {\it Bob}, in the presence of a number of multi-antenna eavesdroppers,
called {\it Eves}.
All the communication links are assumed to undergo slow frequency-flat fading.
The received signals at Bob and Eves may then be modeled as
\begin{subequations} \label{eq:sig_mod}
\begin{align}
y_{b}(t) & = {\bf h}^H {\bf x}(t) + n(t), \\
{\bf y}_{e,k}(t) & = {\bf G}_k^H {\bf x}(t) + {\bf v}_k(t), \quad
k \in {\cal K},
\end{align}
\end{subequations}
respectively,
where ${\cal K} \triangleq \{1,\ldots,K\}$; ${\bf h} \in \mathbb{C}^{N_t}$ is the channel response from
Alice to Bob, with $N_t$ being the number of transmit antennas; ${\bf
G}_k \in \mathbb{C}^{N_t \times N_{e,k}}$ is the channel response
from Alice to $k$th Eve, with $N_{e,k}$ being the number
of receive antennas at $k$th Eve; $K$ is the number of Eves;
$n(t) \sim \mathcal{CN} (0, 1)$ and ${\bf v}_k(t) \sim \mathcal{CN}
({\bf 0}, {\bf I}_{N_{e,k}}) $ are standard additive white complex
Gaussian noises at Bob and $k$th Eve, respectively; ${\bf x}(t)
\in \mathbb{C}^{N_t}$ is the transmit signal vector, which possesses
the following form
\[ {\bf x}(t) = {\bf s}(t) + {\bf z}(t). \]
Here, $\{ {\bf s}(t) \}$ is the coded confidential information intended for Bob,
and  ${\bf z}(t)$ is the noise vector artificially created by Alice to interfere Eves, i.e., the so-called AN.
The confidential signal vector ${\bf s}(t)$ is assumed to follow a complex Gaussian distribution
$\mathcal{CN} ({\bf 0}, {\bf W})$ \cite{Liang_book},
where ${\bf W}$ is the transmit covariance and is to be designed.
For the AN, we assume ${\bf z}(t) \sim \mathcal{CN} ({\bf 0}, {\bm \Sigma})$,
where ${\bm \Sigma}$ is the AN covariance and is again to be designed.
Note that if ${\bf W}$ is chosen such that
${\bf W} = {\bm w} {\bm w}^H$ for some ${\bm w} \in \mathbb{C}^{N_t}$, or equivalently,
${\rm rank}({\bf W}) \leq 1$,
then the transmit strategy for the confidential information is transmit beamforming;
viz., ${\bf s}(t) = {\bm w} s(t)$ where $s(t)$ is a data stream carrying the confidential information.


\subsection{Problem Statement} \label{sec:prob_statement}
Our problem is to design the transmit and AN covariances ${\bf W}, {\bm \Sigma}$ such that maximum information secrecy can be achieved.
Given $({\bf W}, \bm \Sigma)$, an achievable secrecy rate is given by~\cite{Liang2007}
\begin{equation} \label{eq:secrecy_rate}
  R_s = \min_{k \in {\cal K} } \left \{ C_b({\bf W}, \bm \Sigma) - C_{e,k} ({\bf W}, \bm \Sigma) \right\} ,
\end{equation}
where
$C_b({\bf W}, \bm \Sigma)$ and $C_{e,k} ({\bf W}, \bm \Sigma)$ are the mutual information at Bob and Eves, respectively:
\begin{subequations}\label{mutual_inf}
\begin{align}
 C_b({\bf W}, \bm \Sigma)        & =     \log \Big( 1+ \frac{{\bf h}^H {\bf W} {\bf h}}{1 + {\bf h}^H {\bm \Sigma} {\bf h} } \Big),  \label{mutual_inf_b}\\
 C_{e,k} ({\bf W}, \bm \Sigma)   & =     \log \det ( {\bf I} + ({\bf I}
                                        + {\bf G}_k^H {\bm \Sigma} {\bf G}_k)^{-1} {\bf G}_k^H {\bf W} {\bf G}_k ). \label{mutual_inf_c}
\end{align}
\end{subequations}
Note that \eqref{eq:secrecy_rate} is a rate at which perfect secrecy is possible; i.e., Bob
can correctly decode the confidential information at
$R_s$
bits per channel
use, while Eves can retrieve almost nothing about the information.
Readers are referred to the information theoretic security literature, such as~\cite{Liang_book}, for the detail.
Assuming perfect CSI at Alice,
the {\it secrecy-rate maximization (SRM)} design problem is formulated as:
\begin{center}
\fbox{
\parbox{0.8\textwidth}{
\begin{subequations}\label{eq:SRM_main}
  \begin{align}
  R^\star_s  = \max_{{\bf W}\succeq {\bf 0}, {\bm \Sigma}\succeq {\bf 0}}& ~ \min_{k \in {\cal K} } \left\{ C_b({\bf W}, \bm \Sigma) - C_{e,k} ({\bf W}, \bm \Sigma) \right\} \label{eq:SRM_main_a} \\
    {\rm s.t.}& ~ {\rm Tr}({\bf W} + {\bm \Sigma}) \leq P,  \label{eq:SRM_main_b} \\
    & ~ {\rm Tr}({\bm \Phi}_l ( {\bf W} + {\bm \Sigma} ) ) \leq \rho_l, ~ \forall l \in {\cal L}, \label{eq:SRM_main_c}
  \end{align}
\end{subequations}
}
}
\end{center}
where ${\cal L} \triangleq \{1, \ldots, L\}$;
$P >0$ specifies the transmit sum power budget,
and ${\bm \Phi}_l \in \mathbb{H}^{N_t}_+$, $\rho_l \in \mathbb{R}_+, ~\forall l \in {\cal L},$ are given design parameters.
A standard SRM problem has the sum power constraint \eqref{eq:SRM_main_b},
but not \eqref{eq:SRM_main_c}.
In the following, we describe two application-specific
scenarios where
\eqref{eq:SRM_main_c} is necessary.
\begin{itemize}
  \item[1)] {\it Per-antenna power constraints}: \
  In multi-antenna system implementations,
  each antenna is often equipped with its own power amplifier (PA).
  In order to operate within the linear region of each PA,
  one may want to limit the per-antenna peak power~\cite{WEIYU,Gursoy2010}.
  The per-antenna power constraints can be formulated as
  \begin{equation} \label{eq:PAPC}
  \left[{\bf W}+ {\bm \Sigma} \right]_{ll} \leq \rho_l, \quad l=1,\ldots, N_t,
  \end{equation}
  where $\rho_l$ is the power limit of the $l$th antenna.
  The per-antenna power constraints above can be represented by \eqref{eq:SRM_main_c},
  by setting ${\bm \Phi}_l = {\bf e}_l {\bf e}_l^H$, $L= N_t$,
  where ${\bf e}_l$ is the $l$th unit vector
  (i.e., $[ {\bf e}_l ]_i = 1$ for $i= l$ and $[ {\bf e}_l ]_i = 0$ for all $i \neq l$).

\item[2)] {\it Interference temperature constraints}: \
Consider an extension of the secure communication problem setup in Section~\ref{sec:sys_model},
where the system is operating under an interference network scenario.
In such a case,
besides the security concern arising from Eves, Alice also needs to cautiously control her transmission such that no excessive interference will be made to other network users.
Take the spectrum-sharing cognitive radio (CR) network as an example.
Alice and Bob are the secondary transmitter and receiver, respectively.
To limit interference to the primary users, the following interference temperature constraints can be added in the SRM design \cite{yiyangpei2010}:
\begin{equation} \label{eq:ITC}
      {\rm Tr} \left( {\bf R}_l^H ( {\bf W} + {\bm \Sigma} ) {\bf R}_l \right) \leq \rho_l, \quad l=1,\ldots,L,
    \end{equation}
where
${\bf R}_l \in \mathbb{C}^{N_t \times N_{p,l}}$ is the channel response from Alice to $l$th primary user, with $N_{p,l}$ being the number of receive antennas at the $l$th primary user;
$L$ is the number of primary users;
$\rho_l \geq 0$ is the maximal allowable interference level of the $l$th primary user.
The interference temperature constraints \eqref{eq:ITC} can be represented by \eqref{eq:SRM_main_c},
by setting ${\bm \Phi}_l \triangleq {\bf R}_l {\bf R}_l^H$, $l=1,\ldots,L$.
It is worthwhile to note that apart from CR,
the same interference control idea may be applied to multicell interference networks~\cite{Huh10}.
\end{itemize}

\section{An SDP-based Approach to the SRM Problem}\label{sec:perfect_CSI}

In this section, we derive an SDP-based optimization approach to the SRM problem \eqref{eq:SRM_main}.

\subsection{A Tight Relaxation of the SRM Problem \eqref{eq:SRM_main}}\label{sec:tight_relx_SRM}

To start with, we rewrite the SRM problem \eqref{eq:SRM_main} as
\begin{subequations} \label{eq:SRM_reform_1}
  \begin{align}
    R^\star_s =  \max_{{\bf W}\succeq {\bf 0}, {\bm \Sigma}\succeq {\bf 0}, \beta \geq 1}  &~ C_b({\bf W}, \bm \Sigma) - \log \beta \label{eq:SRM_reform_1_a} \\
      {\rm s.t.}  & ~ C_{e,k} ({\bf W} , \bm \Sigma) \leq \log \beta, ~ \forall  k \in {\cal K},  \label{eq:SRM_reform_1_b} \\
  & ~{\rm Tr}({\bf W} + {\bm \Sigma}) \leq P,~ {\rm Tr}({\bm \Phi}_l ( {\bf W} + {\bm \Sigma} ) ) \leq \rho_l, ~\forall l \in {\cal L},\label{eq:SRM_reform_1_c}
  \end{align}
\end{subequations}
where $\beta$ is
a slack variable
introduced to simplify the objective function. Physically, $\log \beta$ can be interpreted as the maximal allowable mutual
information of Eves' links. By adjusting $\beta$, we
can
control the level of mutual information between Alice and Eves,
and consequently, the secrecy rate.
By substituting \eqref{mutual_inf_b} and \eqref{mutual_inf_c} into \eqref{eq:SRM_reform_1},
we express problem \eqref{eq:SRM_reform_1} as
\begin{subequations} \label{eq:SRM_main1}
  \begin{align}
    R^\star_s = \max_{{\bf W} , {\bm \Sigma}, \beta }& ~ \log \left( 1+ \frac{{\bf h}^H
{\bf W} {\bf h}}{1 + {\bf h}^H {\bm \Sigma} {\bf h}
    } \right) - \log \beta \label{eq:SRM_main1_a} \\
    {\rm s.t.}& ~ \log \det\left( {\bf I} + \left({\bf I} + {\bf G}_k^H {\bm \Sigma} {\bf G}_k \right)^{-1} {\bf G}_k^H
    {\bf W} {\bf G}_k \right) \leq \log \beta, ~ \forall k \in {\cal K},  \label{eq:SRM_main1_b} \\
    & ~ {\bf W} \succeq {\bf 0}, ~  {\bm \Sigma}\succeq {\bf 0}, ~ \beta \geq 1, ~{\rm and}~ \eqref{eq:SRM_reform_1_c}~ {\rm satisfied}. \notag
  \end{align}
\end{subequations}
Problem \eqref{eq:SRM_main1} is nonconvex.
In particular,
the most challenging part lies in \eqref{eq:SRM_main1_b},
which can be verified to be nonconvex and may be difficult to deal with.
To circumvent this difficulty, our idea is to
derive
a relatively easy-to-handle inequality
in place of
\eqref{eq:SRM_main1_b}:
\begin{Prop}\label{prop:ineq_relax_key}
The following implication holds
\begin{subequations}\label{eq:fact1}
\begin{align}
\hspace{-5pt}& \log \det\left( {\bf I} + \left({\bf I} + {\bf G}^H {\bm \Sigma}
{\bf G} \right)^{-1} {\bf G}^H
    {\bf W} {\bf G} \right) \leq \log \beta  \label{eq:fact1_a} \\
\hspace{-5pt} \Longrightarrow~ &  (\beta -1) ( {\bf I} + {\bf G}^H {\bm \Sigma} {\bf G}) - {\bf G}^H
    {\bf W} {\bf G} \succeq {\bf 0} \label{eq:fact1_b}
\end{align}
\end{subequations}
for any ${\bf G} \in \mathbb{C}^{N \times M}$,  ${\bf W} \in
\mathbb{H}_{+}^{N}$, and ${\bm \Sigma} \in \mathbb{H}_{+}^{N}$.
Moreover, \eqref{eq:fact1_a} and \eqref{eq:fact1_b} are
equivalent if ${\rm rank}({\bf W}) \leq 1$.
\end{Prop}
The proof of Proposition~\ref{prop:ineq_relax_key} is given in Appendix~\ref{appendix_Prop1}.
From Proposition~\ref{prop:ineq_relax_key}, we note the following:

\begin{Remark}\label{remark:key-relaxation-ineq-1}
  The merit of \eqref{eq:fact1_b} is that for any fixed $\beta$, \eqref{eq:fact1_b} is a convex inequality.
  Specifically, \eqref{eq:fact1_b} is a linear matrix inequality w.r.t. $({\bf W}, {\bm \Sigma})$.
  In comparison,
  in \eqref{eq:fact1_a}, we are confronted with the troublesome matrix inversion and determinant.
\end{Remark}
\begin{Remark}\label{remark:key-relaxation-ineq-2}
  Proposition~\ref{prop:ineq_relax_key} indicates that any $ ( {\bf W}, {\bm
\Sigma}, \beta )$ satisfying \eqref{eq:fact1_a} also satisfies
\eqref{eq:fact1_b}. In other words, \eqref{eq:fact1_b} is a
relaxation of \eqref{eq:fact1_a} in the sense that replacing
\eqref{eq:SRM_main1_b} with \eqref{eq:fact1_b} yields a larger
feasible solution set (or equivalently higher secrecy rate) for the SRM
problem \eqref{eq:SRM_main1}. In addition, such a replacement makes
no difference if ${\rm rank}( {\bf W}) \leq 1$.
\end{Remark}

Now, let us replace \eqref{eq:SRM_main1_b} with \eqref{eq:fact1_b} and consider the subsequent {\it relaxed SRM problem}, which is formulated as follows:
 \begin{subequations} \label{eq:SRM_relax}
\begin{align}
\bar{R}^\star_s   =  \max_{{\bf W} , {\bm \Sigma}, \beta }  & ~ \log \left( \frac{1 + {\bf h}^H ( {\bf W} + {\bm \Sigma} ) {\bf h}} {\beta( 1 + {\bf h}^H {\bm
  \Sigma} {\bf h} ) } \right) \label{eq:SRM_relax_a} \\
   {\rm s.t.} &  ~  (\beta -1) ( {\bf I} + {\bf G}_k^H {\bm \Sigma} {\bf G}_k) \succeq  {\bf G}_k^H
    {\bf W} {\bf G}_k, ~\forall  k \in {\cal K}, \label{eq:SRM_relax_b} \\
   & ~ {\bf W} \succeq {\bf 0}, ~  {\bm \Sigma}\succeq {\bf 0}, ~ \beta \geq 1,~ \eqref{eq:SRM_reform_1_c}~ {\rm satisfied}, \label{eq:SRM_relax_c}
  \end{align}
\end{subequations}
where $\bar{R}^\star_s$ denotes the optimal objective value of problem~\eqref{eq:SRM_relax}.
As discussed in Remark~\ref{remark:key-relaxation-ineq-2},
problem~\eqref{eq:SRM_relax} relaxes the feasible solution set of problem~\eqref{eq:SRM_main1},
and hence has $R^\star_s  \leq   \bar{R}^\star_s$ in general.
Interestingly, we show that $R^\star_s  =  \bar{R}^\star_s$ always holds for problem~\eqref{eq:SRM_relax}.


\begin{Theorem}\label{theorem:perfect-CSI}
Problem~\eqref{eq:SRM_relax} is a tight relaxation to, or an equivalent form of, the SRM problem~\eqref{eq:SRM_main1}.
In particular, there exists an optimal solution $( {\bf W}^\star, {\bm \Sigma}^\star, \beta^\star )$ of problem~\eqref{eq:SRM_relax}, for which ${\rm rank}({\bf W}^\star) \leq 1$;
the solution $( {\bf W}^\star, {\bm \Sigma}^\star, \beta^\star )$ is also an optimal solution of problem \eqref{eq:SRM_main1}, achieving $R^\star_s = \bar{R}^\star_s$.
\end{Theorem}

Theorem~\ref{theorem:perfect-CSI} suggests that
we can equivalently solve the SRM problem \eqref{eq:SRM_main1} by
solving the relaxed (and less difficult) problem~\eqref{eq:SRM_relax}.
The proof of Theorem~\ref{theorem:perfect-CSI} is relegated to Appendix~\ref{appendix_Thm1}. The intuition behind the proof is
the equivalence between \eqref{eq:SRM_relax_b} and \eqref{eq:SRM_main1_b} when
${\rm rank}({\bf W}) \leq 1$,
cf. Proposition~\ref{prop:ineq_relax_key}.
This key observation motivates us to prove the existence of an optimal ${\bf W}^\star$ of problem~\eqref{eq:SRM_relax} that has ${\rm rank}({\bf W}^\star) \leq 1$.
We have the following remarks for Theorem~\ref{theorem:perfect-CSI}:
\begin{Remark}\label{remark-perfect-csi-2}
 Theorem~\ref{theorem:perfect-CSI} implies that
 the SRM problem~\eqref{eq:SRM_main1} admits an optimal ${\bf
  W}^\star$ with ${\rm rank}( {\bf W}^\star) \leq 1$,
  which holds true irrespective of the number of Eves and
  the number of antennas of Eves.
  Physically, it means that
  transmit beamforming is an optimal strategy for the confidential information transmission.

\end{Remark}
\begin{Remark}\label{remark-perfect-csi-3}
  One should note that the solution correspondence between \eqref{eq:SRM_main1} and \eqref{eq:SRM_relax} holds not only at the optimal $\beta^\star$. In fact, the proof of the theorem reveals that given any feasible $\beta$ in \eqref{eq:SRM_main1}, the corresponding optimal $({\bf W}, {\bm \Sigma})$ can be found by solving the relaxation~\eqref{eq:SRM_relax} for the same $\beta$.
\end{Remark}


In Theorem~\ref{theorem:perfect-CSI}, our statement is that a rank-one SRM-optimal ${\bf W}^\star$ exists for problem~\eqref{eq:SRM_relax}.
In fact, the proof of Theorem~\ref{theorem:perfect-CSI} reveals that a rank-one SRM-optimal ${\bf W}^\star$ can always be constructed algorithmically.
Based on the proof, we have the following rank-one solution construction procedure:

\begin{Corollary}\label{corollary:SRM-perfect-csi}
  Suppose that $( \bar{\bf W}^\star, \bar{\bm \Sigma}^\star, {\beta}^\star )$ is an optimal solution returned by solving problem \eqref{eq:SRM_relax}.
  If ${\rm rank}(\bar{\bf W}^\star) \leq 1$,
  then output $( \bar{\bf W}^\star, \bar{\bm \Sigma}^\star, {\beta}^\star )$ as an optimal solution of the SRM problem~\eqref{eq:SRM_main1}.
  Otherwise, solve the following SDP
  \begin{equation} \label{Thm1_Pow_min}
  \begin{aligned}
    ({\bf W}^\star,\bm{\Sigma}^\star) = \arg \min_{{\bf W} \succeq {\bf 0}, {\bm \Sigma} \succeq {\bf 0}} ~ &  {\rm Tr}({\bf W} + {\bm \Sigma})  \\
     {\rm s.t.}  & ~   {\bf h}^H \left ( {\bf W} + \mu {\bm \Sigma} \right) {\bf h} + \mu \geq 0,   \\
& ~ (\beta^\star -1 ) ({\bf I} + {\bf G}_k^H {\bm \Sigma} {\bf G}_k) \succeq {\bf G}_k^H {\bf W} {\bf G}_k,~ \forall  k \in {\cal K}, \\
    &  ~  {\rm Tr}\left({\bm \Phi}_l ({\bf W} + {\bm \Sigma}) \right) \leq \rho_l,~ \forall l \in {\cal L},
  \end{aligned}
\end{equation}
where $\mu = 1- \beta^\star 2^{\bar{R}^\star_s} $,
and output $({\bf W}^\star,\bm{\Sigma}^\star,\beta^\star)$ as an optimal solution of the SRM problem~\eqref{eq:SRM_main1}.
In particular, it must hold true that ${\rm rank}({\bf W}^\star) \leq 1$.
%
%
\end{Corollary}
\noindent
Corollary~\ref{corollary:SRM-perfect-csi} is a direct consequence of the proof of Theorem~\ref{theorem:perfect-CSI}; see Appendix~\ref{appendix_Thm1} for the details.



\subsection{An SDP-based Line Search Method for Relaxed SRM \eqref{eq:SRM_relax}} \label{sec:SDP-search-perfect-CSI}

We now focus on solving the tight SRM relaxation \eqref{eq:SRM_relax} derived in the last subsection.
Problem~\eqref{eq:SRM_relax} can be reformulated as a one-variable optimization problem, which can be efficiently handled by solving a sequence of SDPs. To show this, note that
\begin{equation} \label{eq:upp-bound-beta}
  \beta \leq 1+\frac{{\bf h}^H {\bf W} {\bf h}}{1 + {\bf h}^H {\bf
\Sigma} {\bf h}} \leq 1+{\bf h}^H {\bf W} {\bf h} \leq 1 + P \|{\bf
h}\|^2,
\end{equation}
where the first inequality is due to
\eqref{eq:SRM_relax_a} and $R^\star_s \geq 0$,
and
the third inequality follows from 
the fact that ${\bf h}^H {\bf W} {\bf h} \leq {\rm Tr}({\bf W}) \| {\bf h} \|^2$ for any ${\bf W} \succeq {\bf 0}$ and ${\rm Tr}({\bf W}) \leq P$.
Then we rewrite \eqref{eq:SRM_relax} as
\begin{equation}\label{eq:s1}
  \begin{array}{rl}
\bar{\gamma}^\star = \displaystyle \max_{\alpha} &  \varphi(\alpha) \\
 {\rm s.t.} &  (1+P\|{\bf h}\|^2)^{-1} \leq \alpha \leq 1,
  \end{array}
\end{equation}
where $\log \bar{\gamma}^\star = \bar{R}^\star_s$, $\alpha = 1/\beta$ and
\begin{subequations} \label{eq:s2}
  \begin{align}
   \varphi(\alpha) \triangleq \max_{{\bf W} , {\bm \Sigma} }   & ~  \frac{1 + {\bf h}^H ( {\bf W} + {\bm \Sigma} ) {\bf h}} {\alpha^{-1}( 1 + {\bf h}^H {\bm \Sigma} {\bf h} ) } \label{eq:s2_a} \\
 {\rm s.t.}  & ~ (\alpha^{-1} -1) ( {\bf I} + {\bf G}_k^H {\bm \Sigma} {\bf G}_k) \succeq {\bf G}_k^H {\bf W} {\bf G}_k , ~\forall k \in {\cal K},  \label{eq:s2_b} \\
& ~ {\bf W} \succeq {\bf 0}, ~  {\bm \Sigma}\succeq {\bf 0}, ~{\rm and}~ \eqref{eq:SRM_reform_1_c}~ {\rm satisfied}. \label{eq:s2_c}
  \end{align}
\end{subequations}
The function $\varphi(\alpha)$ does not have a closed form, but is numerically tractable. In particular, \eqref{eq:s2} can be converted to a convex optimization problem. By applying the Charnes-Cooper
transformation~\cite{Charnes1962}, where we introduce a change of
variables
\begin{equation}\label{eq:CCT}
  {\bf W} = {\bf Q}/ {\xi}, \quad {\bm \Sigma} = {\bm \Gamma} / {\xi}, \quad \xi >0,
\end{equation}
we can equivalently express \eqref{eq:s2} as
\begin{subequations} \label{eq:SRM_CCT}
  \begin{align}
  \varphi(\alpha) = \max_{{\bf Q}, {\bm \Gamma}, \xi} & ~ \xi + {\bf h}^H ({\bf Q} + {\bm \Gamma}) {\bf h} \label{eq:SRM_CCT_a} \\
     {\rm s.t.} & ~  \xi + {\bf h}^H {\bm \Gamma } {\bf h} = \alpha, \label{eq:SRM_CCT_b} \\
    & ~ (1- \alpha) ( \xi {\bf I} + {\bf G}_k^H {\bm \Gamma} {\bf G}_k) \succeq \alpha {\bf G}_k^H
    {\bf Q} {\bf G}_k, ~ \forall k \in {\cal K},  \label{eq:SRM_CCT_c} \\
    & ~ {\rm Tr}\left( \bm{\Phi}_l ( {\bf Q} + {\bm \Gamma}) \right) \leq \rho_l \xi, ~\forall l \in {\cal L},  \label{eq:SRM_CCT_d}  \\
    & ~ {\rm Tr}({\bf Q} + {\bm \Gamma}) \leq \xi P,~ {\bf Q}\succeq {\bf 0}, ~ {\bm \Gamma}\succeq {\bf 0}. \label{eq:SRM_CCT_e}
  \end{align}
\end{subequations}
The motivation of the transformation above is that we want to transform the fractional objective function in \eqref{eq:s2_a}, which is quasiconvex but not convex,
to the linear (and convex) objective function in \eqref{eq:SRM_CCT_a}.
Intuitively, the idea is to fix the denominator of \eqref{eq:s2_a}, and that leads to the constraint in \eqref{eq:SRM_CCT_b}.
The proof of the solution equivalence of problems \eqref{eq:s2} and \eqref{eq:SRM_CCT} can be easily obtained by following the argument in \cite{QLI2011}.
The upshot of the transformation above is that
problem~\eqref{eq:SRM_CCT} is a convex SDP,
which can be efficiently and conveniently solved in a globally optimal manner by off-the-shelf conic optimization softwares, e.g. \verb"SeDuMi"~\cite{Sturm1999} and \verb"CVX"~\cite{Grant2011}.
Therefore, the SDP~\eqref{eq:SRM_CCT} provides us with an efficient way to compute $\varphi(\alpha)$ for any fixed $\alpha$. Since $\alpha$ lies in the interval $[(1 + P\|{\bf h}\|^2)^{-1},~1]$, the single-variable optimization problem~\eqref{eq:s1} can be handled by performing a one-dimensional line search over $\alpha$, and choosing the one that leads to the maximum $\varphi(\alpha)$ as an optimal solution of~\eqref{eq:s1}. In the optimization literature, there are many derivative-free search algorithms for solving one-dimensional optimization problems, e.g., compass or coordinate search (cf. \cite{Kolda03,Conn-book09}). In practice, we use either uniform sampling or the golden search \cite{Bertsekas} to obtain a satisfactory solution. Once problem~\eqref{eq:s1} is solved, the solution $({\bf Q}^\star, {\bm \Gamma}^\star, \xi^\star)$ outputted from the SDP~\eqref{eq:SRM_CCT} can be used to
recover ${\bf W}^\star$ and ${\bm \Sigma}^\star$ through the relation \eqref{eq:CCT}. Note that an additional rank-one solution construction procedure may be needed depending on the rank of ${\bf W}^\star$. This can be done by further solving the SDP~\eqref{Thm1_Pow_min},
cf. Corollary~\ref{corollary:SRM-perfect-csi}.


Summarizing the development in this section,
we have presented an SDP-based optimization approach to the SRM problem~\eqref{eq:SRM_main}.
The approach is based on solving problem~\eqref{eq:SRM_relax}, which is equivalent to problem~\eqref{eq:SRM_main} as our analysis has revealed.
To solve problem~\eqref{eq:SRM_relax},
we have proposed an SDP-based line search formulation in Section~\ref{sec:SDP-search-perfect-CSI}.
In addition to design optimization,
our development has shown that the transmit beamforming strategy is SRM-optimal for the transmission of confidential information.

\section{Extension to Robust SRM}\label{sec:partial_CSI}

Our next endeavor is to extend the optimization approach developed in the last section to an imperfect CSI case,
where Alice has incomplete knowledge of Eves' CSI.
Specifically, we consider a worst-case robust SRM (WCR-SRM) formulation under norm-bounded CSI uncertainties,
and derive an SDP-based solution approach to the problem.
We will also illustrate how the developed WCR-SRM design solution can be used to provide a safe approximation to an even more challenging design, namely, the outage-constrained robust design (OCR-SRM) under Gaussian distributed CSI uncertainties.



\vspace{-5pt}
\subsection{The Worst-Case Robust SRM Problem} \label{sec:wc-srm}

We consider the same problem setup as in Section~\ref{sec:prob_set},
with the additional assumption that Alice has imperfect CSI on Eves' links.
To put into context, let
\begin{equation}\label{WC_CSI_model}
  {\bf G}_k = \bar{\bf G}_k + \Delta {\bf G}_k,\quad k=1,\ldots,K,
\end{equation}
where ${\bf G}_k$ is the {\it actual} channel response from Alice to the $k$th Eve, as before;
$\bar{\bf G}_k$ is Alice's {\it presumed} value of ${\bf G}_k$;
$\Delta {\bf G}_k$ represents the associated CSI error.
In the WCR-SRM formulation, we assume that $\Delta {\bf G}_k$ are deterministic and bounded, satisfying~\cite{JHuang12,Wolf10}
\[ \| \Delta {\bf G}_k \|_F \leq \varepsilon_k, ~k=1,\ldots,K \]
for some $\varepsilon_k>0,~k=1,\ldots,K$.
The WCR-SRM design problem is formulated as
\begin{center}
\fbox{
\parbox{0.85\textwidth}{
\begin{subequations}\label{eq:WC_SRM_main}
  \begin{align}
     \hspace{-10pt} {R}^\star_s  =   \max_{{\bf W}\succeq {\bf 0} , {\bm \Sigma}\succeq {\bf 0} } &  \min_{k \in {\cal K}}  \left\{ C_b ({\bf W}, {\bm \Sigma}) - C_{e,k}^{\rm worst} ({\bf W}, {\bm \Sigma}) \right\} \label{eq:WC_SRM_main_a} \\
        {\rm s.t.} & ~{\rm Tr}({\bf W} + {\bm \Sigma}) \leq P, \label{eq:WC_SRM_main_b} \\
        & ~ {\rm Tr} \left({\bm \Phi}_l ( {\bf W} + {\bm \Sigma} ) \right) \leq \rho_l,~ \forall l \in {\cal L},\label{eq:WC_SRM_main_c}
  \end{align}
\end{subequations}
}
}
\end{center}
where
we recall that $C_b ({\bf W}, {\bm \Sigma} ) =
\log \left(1 + \frac{{\bf h}^H {\bf W} {\bf h}}{1 + {\bf h}^H {\bm \Sigma} {\bf h} } \right)$ is Bob's mutual information, and
\begin{align*}
C_{e,k}^{\rm worst} ({\bf W}, {\bm \Sigma}) &  \triangleq  \max_{{\bf G}_k \in \mathcal{B}_k} \log \det \big( {\bf I} + \big({\bf I} + {\bf G}_k^H {\bm \Sigma} {\bf G}_k \big)^{-1} {\bf G}_k^H
{\bf W} {\bf G}_k \big), \\
\mathcal{B}_k &  \triangleq \left\{ {\bf G}_k ~|~ {\bf G}_k = \bar{\bf G}_k + \Delta {\bf G}_k,~ \| \Delta {\bf G}_k \|_F \leq \varepsilon_k \right\}, \ \forall k \in {\cal K}
\end{align*}
is the $k$th-Eve's worst-case, or largest possible, mutual information among the set of all admissible CSIs $\mathcal{B}_k$.
Problem~\eqref{eq:WC_SRM_main} is a safe design---under the optimal design $({\bf W},{\bm \Sigma})$ of problem~\eqref{eq:WC_SRM_main}, the actual secrecy rate w.r.t. the true ${\bf G}_k$'s, albeit uncertain, 
must not lie below the optimal worst-case secrecy rate ${R}^\star_s$. 

Our WCR-SRM optimization approach is derived as follows.
We rewrite \eqref{eq:WC_SRM_main} as
\begin{subequations}\label{eq:WC-SRM-reform}
  \begin{align}
{R}^\star_s   =  \max_{{\bf W}\succeq {\bf 0},  {\bm \Sigma} \succeq {\bf 0}, \beta \geq 1 }  & ~  \log \big(\frac{1 + {\bf h}^H ( {\bf W} + {\bm \Sigma} ) {\bf h}}{ 1 + {\bf h}^H {\bm \Sigma} {\bf h}  } \big) - \log \beta \label{eq:WC-SRM-reform-a} \\
       {\rm s.t.} &¡¡~   \log \det\big( {\bf I} + \big({\bf I} + {\bf G}_k^H {\bm \Sigma} {\bf G}_k \big)^{-1} {\bf G}_k^H {\bf W} {\bf G}_k \big) \leq \log \beta, ~ \forall\, {\bf G}_k \in \mathcal{B}_k, \ k \in {\cal K},  \label{eq:WC-SRM-reform-b} \\
         &~ \eqref{eq:WC_SRM_main_b}-\eqref{eq:WC_SRM_main_c}~{\rm satisfied}. \label{eq:WC-SRM-reform-c}
  \end{align}
\end{subequations}
Note that in \eqref{eq:WC-SRM-reform-b}, there are infinitely many inequalities
w.r.t. ${\bf G}_k$ to satisfy;
this makes the WCR-SRM problem more challenging to solve than the SRM.
Let us set aside the infinitely many inequalities issue for the moment.
By using Proposition~\ref{prop:ineq_relax_key},
which has played a key role in solving SRM in the last section,
we have the following relaxation for \eqref{eq:WC-SRM-reform-b}:
\begin{subequations}\label{eq:semi-inf2mat-ineq}
  \begin{align}
& \log \det\big( {\bf I} + \big({\bf I} + {\bf
G}_k^H {\bm \Sigma} {\bf G}_k \big)^{-1} {\bf G}_k^H
    {\bf W} {\bf G}_k \big) \leq \log \beta, ~\forall\, {\bf G}_k \in \mathcal{B}_k, \label{eq:semi-inf2mat-ineq-a} \\
 \Longrightarrow &  ( \beta -1) ( {\bf I} +
{\bf G}_k^H {\bm \Sigma} {\bf G}_k) \succeq  {\bf G}_k^H
    {\bf W} {\bf G}_k ,\,\forall\,{\bf G}_k \in \mathcal{B}_k,  \label{eq:semi-inf2mat-ineq-b}
  \end{align}
\end{subequations}
for $k=1,\ldots,K$.
Moreover, \eqref{eq:semi-inf2mat-ineq-a} and \eqref{eq:semi-inf2mat-ineq-b} become equivalent if ${\rm rank}({\bf W}) \leq 1$ (again, by Proposition~\ref{prop:ineq_relax_key}).
Eq.~\eqref{eq:semi-inf2mat-ineq-b} corresponds to an infinite number of quadratic matrix inequalities w.r.t. ${\bf G}_k$.
While \eqref{eq:semi-inf2mat-ineq-b} is less complex than \eqref{eq:semi-inf2mat-ineq-a},
we still need to find an efficient way to manage the infinitely many inequalities in \eqref{eq:semi-inf2mat-ineq-b}.
It turns out that the latter is possible, by employing an advanced matrix inequality result in the optimization literature.
\begin{Lemma}[Luo-Sturm-Zhang~\cite{luo04}] \label{S-lemma}
Let $f({\bf X}) = {\bf X}^H {\bf A} {\bf X} + {\bf X}^H {\bf B} + {\bf B}^H {\bf X} + {\bf C}$, and ${\bf D} \succeq {\bf 0}$. The following equivalence holds:
\begin{equation}\label{eq:lemma_QMI}
\begin{aligned}
& f({\bf X}) \succeq {\bf 0}, ~ \forall ~{\bf X} \in \{ {\bf X}~|~{\rm Tr}( {\bf D} {\bf X}
{\bf X}^H ) \leq 1 \}, \\
  \Longleftrightarrow &  \begin{bmatrix}
{\bf C} & {\bf B}^H \\
{\bf B} & {\bf A}
  \end{bmatrix}
  - t \begin{bmatrix}
    {\bf I} & {\bf 0}\\
    {\bf 0} & -{\bf D}
  \end{bmatrix}
  \succeq {\bf 0}, ~{\rm for ~some~ } t \geq 0.
  \end{aligned}
\end{equation}
\end{Lemma}

By applying Lemma~\ref{S-lemma} to \eqref{eq:semi-inf2mat-ineq-b}, we establish the following key result:
\begin{Prop} \label{prop:wc-s-lemma-implication}
  The following implication holds
  \begin{subequations}\label{eq:mat_ineq_relax}
  \begin{align}
   & \log \det\big( {\bf I} + \big({\bf I} + {\bf
G}_k^H {\bm \Sigma} {\bf G}_k \big)^{-1} {\bf G}_k^H
    {\bf W} {\bf G}_k \big)  \leq \log \beta, ~\forall \,
 {\bf G}_k \in \mathcal{B}_k,  \label{eq:mat_ineq_relax_a} \\
\Longrightarrow  &~  {\bf T}_k (\beta, {\bf W}, {\bm \Sigma},
t_k ) \succeq {\bf 0}, ~{\rm for ~ some ~} t_k \geq 0,  \label{eq:mat_ineq_relax_b}
  \end{align}
\end{subequations}
for any $k \in {\cal K}$, where
\begin{equation} \label{eq:Tk}
{\bf T}_k (\beta, {\bf W}, {\bm \Sigma},
t_k ) = \begin{bmatrix} ( \beta- 1 - t_k) {\bf I} + \bar{\bf
G}_k^H \left( (\beta - 1 ) {\bm \Sigma} - {\bf W} \right)
\bar{\bf G}_k  & \bar{\bf G}_k^H  \left( (\beta -1
) {\bm \Sigma} - {\bf W} \right) \\
  \left( (\beta -1 ) {\bm \Sigma} - {\bf W} \right) \bar{\bf G}_k & (\beta -1)
{\bm \Sigma} - {\bf W} + \frac{t_k}{\varepsilon_k^2} {\bf I}
  \end{bmatrix}.
\end{equation}
Moreover, \eqref{eq:mat_ineq_relax_a} and \eqref{eq:mat_ineq_relax_b} are equivalent if ${\rm rank}({\bf W}) \leq 1$.
\end{Prop}

\noindent {\it Proof:} \
Following \eqref{eq:semi-inf2mat-ineq}, it suffices to show that for each $k$, \eqref{eq:semi-inf2mat-ineq-b} is equivalent to \eqref{eq:mat_ineq_relax_b}.
Eq.~\eqref{eq:semi-inf2mat-ineq-b} can be represented by the left-hand side of \eqref{eq:lemma_QMI}.
Specifically, we substitute \eqref{WC_CSI_model} into \eqref{eq:semi-inf2mat-ineq-b}, and then set ${\bf X} = \Delta {\bf G}_k$, ${\bf A} = (\beta -1) {\bm
\Sigma - {\bf W}}$, ${\bf B} = \left( (\beta -1) {\bm \Sigma -
{\bf W}}\right) \bar{\bf G}_k$, ${\bf C} = \bar{\bf G}_k^H \left(
(\beta -1) {\bm \Sigma - {\bf W}}\right) \bar{\bf G}_k +
(\beta -1) {\bf I}$ and ${\bf D} = \varepsilon_k^{-2} {\bf
I}$.
By the right-hand side of \eqref{eq:lemma_QMI}, we obtain \eqref{eq:mat_ineq_relax_b} as an equivalent form of \eqref{eq:semi-inf2mat-ineq-b}.
\hfill $\blacksquare$

The upshot of the implication in Proposition~\ref{prop:wc-s-lemma-implication} is that
fixing $\beta$, \eqref{eq:mat_ineq_relax_b} is a {\it single} linear matrix inequality w.r.t. $({\bf W}, {\bm \Sigma}, t_k)$ (rather than infinitely many),
and can be efficiently handled by convex conic optimization.
Therefore, we replace \eqref{eq:WC-SRM-reform-b} by \eqref{eq:mat_ineq_relax_b} to obtain a relaxation of \eqref{eq:WC-SRM-reform}, given as follows:
\begin{equation}\label{eq:WC-SRM-relax}
  \begin{aligned}
       \bar{R}^\star_s =  \max_{{\bf W}\succeq {\bf 0}  , {\bm \Sigma}\succeq {\bf 0} , \{t_k\}, \beta \geq 1 } & ~   \log \Big(\frac{1 + {\bf h}^H ( {\bf W} + {\bm \Sigma} ) {\bf h}}{\beta ( 1 + {\bf h}^H {\bm \Sigma} {\bf h} ) } \Big)  \\
       {\rm s.t.} & ~   {\bf T}_k (\beta, {\bf W}, {\bm \Sigma}, t_k ) \succeq {\bf 0},~ t_k \geq 0, \, \forall k \in {\cal K}  \\
       & ~  \eqref{eq:WC_SRM_main_b}-\eqref{eq:WC_SRM_main_c}~{\rm satisfied},
  \end{aligned}
\end{equation}
where $\bar{R}^\star_s$ is the optimal objective value of \eqref{eq:WC-SRM-relax} and we have ${R}^\star_s \leq \bar{R}^\star_s$.
Now, a crucial question is whether \eqref{eq:WC-SRM-relax} is a tight relaxation of the WCR-SRM problem \eqref{eq:WC-SRM-reform}.
Remarkably, we prove that the answer is yes.
\begin{Theorem}\label{theorem_WC}
  Problem \eqref{eq:WC-SRM-relax} is a tight relaxation to, or an equivalent formulation of, the WCR-SRM problem  \eqref{eq:WC-SRM-reform}.
  In particular, there exists an optimal solution $( {\bf W}^\star, {\bm \Sigma}^\star, \beta^\star )$ of problem
  \eqref{eq:WC-SRM-relax}, for which ${\rm rank}({\bf W}^\star) \leq 1$;
  the solution $( {\bf W}^\star, {\bm \Sigma}^\star, \beta^\star )$ is also an optimal solution of
  \eqref{eq:WC-SRM-reform}, achieving ${R}^\star_s = \bar{R}^\star_s$.
\end{Theorem}
\begin{Corollary}\label{corollary:wc-srm}
Suppose that $( \bar{\bf W}^\star, \bar{\bm \Sigma}^\star, {\beta}^\star )$ is an optimal solution returned by solving~\eqref{eq:WC-SRM-relax}.
If ${\rm rank} (\bar{\bf W}^\star) \leq 1$, then output $( \bar{\bf W}^\star, \bar{\bm \Sigma}^\star, {\beta}^\star )$ as an optimal solution of the WCR-SRM problem~\eqref{eq:WC-SRM-reform}.
Otherwise, solve the following SDP
\begin{equation} \label{eq:wc_srm_pow_min}
  \begin{array}{rl}
   ( {\bf W}^\star, {\bm \Sigma}^\star, \{ t_k^\star \} ) \displaystyle =  \arg \min_{{\bf W} , {\bm \Sigma}, \{ t_k \} } &  {\rm Tr}({\bf W} + {\bm \Sigma})  \\
    {\rm s.t.} &   {\bf h }^H ( {\bf W} + \mu {\bm \Sigma} ) {\bf h} + \mu \geq 0, \\
    &  {\bf T}_k (\beta^\star, {\bf W}, {\bm \Sigma}, t_k) \succeq {\bf 0}, ~ t_k \geq 0,~ \forall  k \in {\cal K}, \\
    & {\rm Tr}({\bm \Phi}_l ({\bf W} + {\bm \Sigma})) \leq \rho_l, ~\forall l \in {\cal L}, \\
    & {\bf W} \succeq {\bf 0},~ {\bm \Sigma} \succeq {\bf 0},
  \end{array}
\end{equation}
where $\mu = 1- 2^{\bar{R}^\star_s} \beta^\star$,
and output $( {\bf W}^\star, {\bm \Sigma}^\star, {\beta}^\star )$ as an optimal solution of the WCR-SRM problem~\eqref{eq:WC-SRM-reform}.
In particular, it must hold true that ${\rm rank}( {\bf W}^\star ) \leq 1$.
\end{Corollary}

Theorem~\ref{theorem_WC} and Corollary~\ref{corollary:wc-srm} can be seen as a generalization of their perfect-CSI counterpart in Theorem~\ref{theorem:perfect-CSI} and Corollary~\ref{corollary:SRM-perfect-csi}, respectively.
The proof of the former is more difficult to obtain than that of the latter, owing to the complicated structure of ${\bf T}_k (\beta, {\bf W}, {\bm \Sigma}, t_k )$ [see \eqref{eq:Tk}].
We relegate the proof to Appendix~\ref{appendix:theorem-wc}.

Since we have identified that \eqref{eq:WC-SRM-relax} is a tight relaxation of the WCR-SRM problem \eqref{eq:WC-SRM-reform},
our last step is to solve \eqref{eq:WC-SRM-relax}.
Problem~\eqref{eq:WC-SRM-relax} can be handled by using the same SDP-based line search method developed in the last section.
For conciseness, here we only point out several key steps.
We reexpress \eqref{eq:WC-SRM-relax} in the form of the one-dimensional problem in \eqref{eq:s1},
where $\varphi(\alpha)$ is now given by
\setlength \arraycolsep{0pt}
\begin{equation} \label{eq:s2_WCR}
  \begin{aligned}
  \varphi(\alpha) =   \max_{{\bf W} \succeq {\bf 0}, {\bm \Sigma} \succeq {\bf 0}, \{ t_k \} }  &  ~   \frac{1 + {\bf h}^H ( {\bf W} + {\bm \Sigma} ) {\bf h}} {\alpha^{-1}( 1 + {\bf h}^H {\bm \Sigma} {\bf h} ) }  \\
 {\rm s.t.} & ~  {\bf T}_k (\alpha^{-1}, {\bf W}, {\bm \Sigma}, t_k ) \succeq {\bf 0}, ~ t_k \geq 0, \,\forall k \in {\cal K}, \\
&~  {\rm Tr}({\bf W} + {\bm \Sigma}) \leq P, ~{\rm Tr} \left({\bm \Phi}_l ( {\bf W} + {\bm \Sigma} ) \right) \leq \rho_l, \, \forall l \in {\cal L}.
  \end{aligned}
\end{equation}
By a change of variables
${\bf W} = {\bf Q}/ {\xi}, {\bm \Sigma} = {\bm \Gamma} / {\xi}, t_k = \lambda_k / {\xi}, ~\xi >0,$
and using the Charnes-Cooper transformation,
we show that \eqref{eq:s2_WCR} can be converted to a convex SDP
\begin{equation}\label{eq:wc-srm-relax-cct2}
  \begin{aligned}
\max_{{\bf Q} \succeq {\bf 0}, {\bm \Gamma} \succeq {\bf 0}, \xi, \{\lambda_k\}}  & ~ \xi + {\bf h}^H
( {\bf Q} + {\bm \Gamma } ) {\bf h}  \\
  {\rm s.t.}  & ~  \xi + {\bf h}^H {\bm \Gamma}  {\bf h}  = \alpha , \\
&  ~ \bar{\bf T}_k ( {\bf Q}, {\bm
\Gamma} , \alpha, \xi, \lambda_k) \succeq
{\bf 0}, ~ \lambda_k \geq 0, ~ \forall k \in {\cal K}, \\
                    &  ~ {\rm Tr}({\bf Q} + {\bm \Gamma}) \leq P \xi, ~ {\rm Tr}\left({\bm \Phi}_l ({\bf Q} + {\bm \Gamma})\right) \leq \rho_l \xi, ~\forall l \in {\cal L},
  \end{aligned}
\end{equation}
where
\[
\bar{\bf T}_k ( {\bf Q}, {\bm
\Gamma} , \alpha, \xi, \lambda_k) =
\begin{bmatrix}
   ( \xi - \alpha \xi - \alpha \lambda_k )  {\bf I} + \bar{\bf
G}_k^H \left( (1 - \alpha ) {\bm \Gamma} - \alpha {\bf Q} \right) \bar{\bf
G}_k & ~~\bar{\bf G}_k^H  \left( ( 1 - \alpha
) {\bm \Gamma} - \alpha {\bf Q} \right) \\
  \left( (1 - \alpha ) {\bm \Gamma} - \alpha {\bf Q} \right) \bar{\bf G}_k & (1 - \alpha)
{\bm \Gamma} - \alpha {\bf Q} + \frac{\alpha \lambda_k}{\varepsilon_k^2} {\bf I}
\end{bmatrix}.
\]
Hence, for a fixed $\alpha$, $\varphi(\alpha)$ can be computed by solving the SDP \eqref{eq:wc-srm-relax-cct2}.
Problem~\eqref{eq:WC-SRM-relax} is then handled by applying a line search on $\varphi(\alpha)$ w.r.t. $\alpha$;
the procedure is the same as that described in Section~\ref{sec:SDP-search-perfect-CSI}.

Summarizing, in this section we have tackled the WCR-SRM problem~\eqref{eq:WC_SRM_main} by deriving a tight relaxation.
The tight WCR-SRM relaxation, given in \eqref{eq:WC-SRM-relax}, can be handled by an SDP-based line search procedure.
Furthermore, it is worthwhile to mention that by Theorem~\ref{theorem_WC},
transmit beamforming is still an optimal  confidential information transmission strategy for the WCR-SRM formulation.

\begin{Remark}
One may be curious to know whether the AN-aided WCR-SRM solution method developed above can be extended to the scenario where Bob's channel is also imperfectly known.
In fact, this has been considered for the no-AN case~\cite{QLI2011}.
For the AN-aided case here, such an extension is possible; the idea is to follow the same derivations as above and \cite{QLI2011}.
It can be shown that the problem can once again be handled via one-dimensional line search, but the line search involves solving a sequence of factional SDPs, which is quasiconvex and is computationally more expensive to solve.
As a future direction, it will be interesting to study efficient methods for handling this problem.
\end{Remark}

\subsection{The Outage-Constrained Robust SRM Problem} \label{sec:oc-srm}

The WCR-SRM formulation in the previous subsection is an absolutely safe design under bounded CSI uncertainties.
In this subsection, we consider an alternative robust formulation where the CSI errors $\Delta {\bf G}_k$ are random and follow certain distribution.
Specifically, we employ the popular i.i.d. complex Gaussian CSI error model
\begin{equation} \label{eq:oc-srm-model}
  [\Delta {\bf G}_k]_{m,n} \sim \mathcal{CN} (0,\sigma_k^2), \quad \forall m,n,~k=1,\ldots,K.
\end{equation}
Moreover, $\Delta {\bf G}_k$ is assumed to be independent of $\Delta {\bf G}_l$ for any $k \neq l$.
In this setup, since $\Delta {\bf G}_k$ are unbounded,
it may not be possible to deliver an absolutely safe design.
However, one can adopt a $(1- \delta) \%$ safe design, for some outage probability specification $\delta$~\cite{Bloch08,Gerbracht12}.
Consider the following outage-constrained robust SRM formulation:
\begin{center}
\fbox{
\parbox{0.85\textwidth}{
\begin{subequations}\label{eq:OC-SRM}
 \begin{align}
  \max_{{\bf W}, {\bm \Sigma}, R_s}& ~ R_s  \label{eq:OC-SRM-a} \\
      {\rm s.t.}     &     ~ {\rm Pr}_{\{ \Delta {\bf G}_k\}_{k=1}^K} \big\{ C_b({\bf W}, {\bm \Sigma}) - \max_{k \in {\cal K}} C_{e,k}({\bf W}, {\bm \Sigma}) \geq R_s \big \} \geq 1 - \delta, \label{eq:OC-SRM-b}\\
                            & ~   {\rm Tr}\left(\bm{\Phi}_l ({\bf W} + {\bm \Sigma}) \right) \leq \rho_l,~\forall l \in {\cal L}, \\
                             & ~ {\rm Tr}({\bf W} + {\bm \Sigma}) \leq P,~ {\bf W}\succeq {\bf 0}, ~{\bm \Sigma}\succeq {\bf 0}, \label{eq:OC-SRM-c}
  \end{align}
\end{subequations}
}
}
\end{center}
where $C_b({\bf W}, {\bm \Sigma})$ and $C_{e,k}({\bf W}, {\bm \Sigma})$ have been defined in \eqref{mutual_inf}; $0< \delta < 0.5$ is a given parameter specifying the maximum tolerable secrecy outage probability, i.e., the probability of the achievable secrecy rate falling below $R_s$~\cite{Bloch08}.
Therefore, the goal of \eqref{eq:OC-SRM} is to maximize the $\delta\%$-outage secrecy rate.

The OCR-SRM problem~\eqref{eq:OC-SRM} is very challenging to solve.
The main obstacle lies in the probabilistic constraint~\eqref{eq:OC-SRM-b},
which is unlikely to have a tractable closed-form expression except for some special cases, e.g., when there is only one Eve with a single antenna and AN is restricted to lie on the nullspace of ${\bf h}$~\cite{Gerbracht12}.
As a compromise,
we consider an approximation of \eqref{eq:OC-SRM} based on the previous WCR-SRM design, which we have solved. We summarize our main result in the following proposition:

\begin{Prop}\label{prop:ocr-srm}
  Consider the WCR-SRM problem~\eqref{eq:WC_SRM_main}. Suppose that the CSI error radii $\varepsilon_k$ are chosen as
  \begin{equation} \label{eq:prop-ocr-srm}
    \varepsilon_k = \sqrt{\frac{\sigma_k^2}{2}
F_{\chi_{2N_tN_{e,k}}^2}^{-1} ( (1-\delta)^{1/K} ) }, ~~ k=1,\ldots,K ,
  \end{equation}
where $F_{\chi_{2N_tN_{e,k}}^2}^{-1} (\cdot)$ denotes the inverse cumulative
distribution function of a Chi-square random variable with
$2N_tN_{e,k}$ degrees of freedom. Then, problem~\eqref{eq:WC_SRM_main} is a safe approximation to the OCR-SRM problem~\eqref{eq:OC-SRM}, in the sense that
every feasible point of problem~\eqref{eq:WC_SRM_main} is also feasible to problem~\eqref{eq:OC-SRM},
and thus satisfies the secrecy rate satisfaction probability constraint~\eqref{eq:OC-SRM-b}.
\end{Prop}

The insight behind Proposition~\ref{prop:ocr-srm} is that the robust nature of WCR-SRM should lead to certain secrecy rate satisfaction probability values;
especially, the larger $\varepsilon_k$ are, the more conservative the design would be and the higher the secrecy rate satisfaction probability should be.
Eq.~\eqref{eq:prop-ocr-srm} is essentially a sufficient condition under which any feasible point of the WCR-SRM problem satisfies the OCR-SRM probability constraint in \eqref{eq:OC-SRM-b}.

\noindent{\it Proof of Proposition~\ref{prop:ocr-srm}:} \
By noting the independence between ${\bf G}_k$ and ${\bf G}_l,~\forall k \neq l$, we have
\begin{subequations}\label{OC-SRM:prob_decouple}
\begin{align}
  \eqref{eq:OC-SRM-b} & \Longleftrightarrow  \prod_{k=1}^{K} {\rm Pr }_{\Delta{\bf G}_k} \left\{ C_b({\bf W}, {\bm \Sigma}) - C_{e,k}({\bf W}, {\bm \Sigma}) \geq R_s  \right\} \geq  1 - \delta, \label{OC-SRM:prob_decouple_a} \\
  & \Longleftarrow  {\rm Pr }_{\Delta {\bf G}_k} \left\{ C_b({\bf W}, {\bm \Sigma}) - C_{e,k}({\bf W}, {\bm \Sigma}) \geq R_s  \right\} \geq  1 - \bar{\delta},~ \forall k \label{OC-SRM:prob_decouple_b}
\end{align}
\end{subequations}
where $\bar{\delta} = 1 - (1-\delta)^{1/K}$.
Our next step is to derive a safe approximation to \eqref{OC-SRM:prob_decouple_b}.
This is done by exploiting the following implication:
\begin{equation}\label{eq:OC-SRM-SB-approx}
\begin{aligned}
& C_b({\bf W}, {\bm \Sigma}) - C_{e,k}({\bf W}, {\bm \Sigma}) \geq R_s, ~\forall \ \|\Delta {\bf G}_k \|_F^2 \leq \varepsilon_k^2 \\
\Longrightarrow & {\rm Pr }_{\Delta {\bf G}_k} \left\{ C_b({\bf W}, {\bm \Sigma}) - C_{e,k}({\bf W}, {\bm \Sigma}) \geq R_s  \right\} \geq  1 - \bar{\delta},
\end{aligned}
\end{equation}
where the uncertainty radius $\varepsilon_k$ is chosen to satisfy ${\rm Pr }_{\Delta {\bf G}_k} \{ \| \Delta {\bf G}_k \|_F^2 \leq \varepsilon_k^2 \} = 1 - \bar{\delta} $. The above implication can be deduced using an intuitive argument called sphere bounding. Specifically, in the left-hand side of \eqref{eq:OC-SRM-SB-approx}, we set a spherical boundary $\varepsilon_k$ to the unbounded random error $\Delta {\bf G}_k$, such that $1 - \bar{\delta}$ portion of $\Delta {\bf G}_k$'s realizations lies in the sphere, and meanwhile, for all of these $\Delta {\bf G}_k$ in the sphere, the worst secrecy rate is ensured no less than $R_s$. For such a choice of $\varepsilon_k$, the right-hand side of \eqref{eq:OC-SRM-SB-approx} apparently holds. In addition, it follows from \eqref{eq:oc-srm-model} that $\varepsilon_k$ can be explicitly calculated by \eqref{eq:prop-ocr-srm}.
Readers are referred to the literature, such as~\cite{KYWang11}, for more complete descriptions of the sphere bounding method.
Now, by applying the implications \eqref{OC-SRM:prob_decouple} and \eqref{eq:OC-SRM-SB-approx} to the probabilistic constraint~\eqref{eq:OC-SRM-b},
we obtain a restriction of, or safe approximation to, the OCR-SRM problem~\eqref{eq:OC-SRM}:
\begin{subequations} \label{eq:WC_SRM_main:prop:ocr-srm}
  \begin{align}
         \max_{{\bf W}, {\bm \Sigma}, R_s} & ~R_s   \\
          {\rm s.t.}  & ~ C_b({\bf W}, {\bm \Sigma}) - C_{e,k}({\bf W}, {\bm \Sigma}) \geq R_s, \forall \ \|\Delta {\bf G}_k \|_F^2 \leq \varepsilon_k^2, ~ k =1, \ldots, K,  \\
                    & ~ {\rm Tr}\left(\bm{\Phi}_l ({\bf W} + {\bm \Sigma}) \right) \leq \rho_l,~\forall l \in {\cal L}, \\
                    &~ {\rm Tr}({\bf W} + {\bm \Sigma}) \leq P,~ {\bf W}\succeq {\bf 0}, ~ {\bm \Sigma}\succeq {\bf 0}.
  \end{align}
\end{subequations}
Note that any feasible point of \eqref{eq:WC_SRM_main:prop:ocr-srm} is also feasible to the OCR-SRM problem~\eqref{eq:OC-SRM}, by \eqref{OC-SRM:prob_decouple} and \eqref{eq:OC-SRM-SB-approx}.
It can be easily seen that problem~\eqref{eq:WC_SRM_main:prop:ocr-srm} is equivalent to the WCR-SRM problem~\eqref{eq:WC_SRM_main}.
\hfill $\blacksquare$

\section{Simulation Results}
In this section, we use Monte Carlo simulations to demonstrate the performance of the AN-aided SRM design obtained by the proposed SDP-based optimization technique.

\subsection{Example 1: Secrecy Rate Performance under the Sum Power Constraint}\label{sec:sim_perfect_CSI}

In this example, our goal is to illustrate the secrecy rate performance of the standard AN-aided SRM design; i.e., problem~\eqref{eq:SRM_main} with the sum power constraint~\eqref{eq:SRM_main_b} only.
We compare the performance of the proposed AN-aided SRM design (Section~\ref{sec:perfect_CSI}) with that of two existing designs, namely, the isotropic AN design~\cite{XYZHOU}  and the optimal no-AN SRM design~\cite{QLI2011}.
In the isotropic AN design, the transmit and AN covariances are given by~\cite{XYZHOU}
\begin{equation}\label{eq:isotropic-AN}
  {\bf W}_{\rm iso-AN} = \frac{P}{2 \| {\bf h} \|^2 } {\bf h} {\bf h}^H, \quad \quad {\bm \Sigma}_{\rm iso-AN} = \frac{P}{2} \frac{{\bm \Pi}_{\bf h}^\perp}{\| {\bm \Pi}_{\bf h}^\perp \|_F^2},
\end{equation}
where ${\bm \Pi}_{\bf h}^\perp = {\bf I} - {\bf h} {\bf h}^H / \| {\bf h} \|^2$ denotes the orthogonal complement projector of ${\bf h}$.
In words, the isotropic AN design uses half of the transmit power to transmit the confidential information via straight beamforming,
and uses the other half to transmit AN on the nullspace of ${\bf h}$ isotropically.
The no-AN SRM design is the solution of problem~\eqref{eq:SRM_main}, but with ${\bm \Sigma}$ prefixing as ${\bf 0}$.
The no-AN SRM design can be obtained by solving one SDP, as shown in our previous work~\cite{QLI2011} (also \cite{QLI_ISPACS10}).

The simulation settings are as follows.
The number of transmit antennas at Alice is $N_t = 5$.
The number of receive antennas at Eves is $N_{e,k} = 3$ for all $k$.
In each simulation trial, Bob and Eves' channels are randomly generated following an i.i.d. complex Gaussian distribution with zero mean and unit variance.

\begin{figure}[htp!]
\centerline{\resizebox{.45\textwidth}{!}{\includegraphics{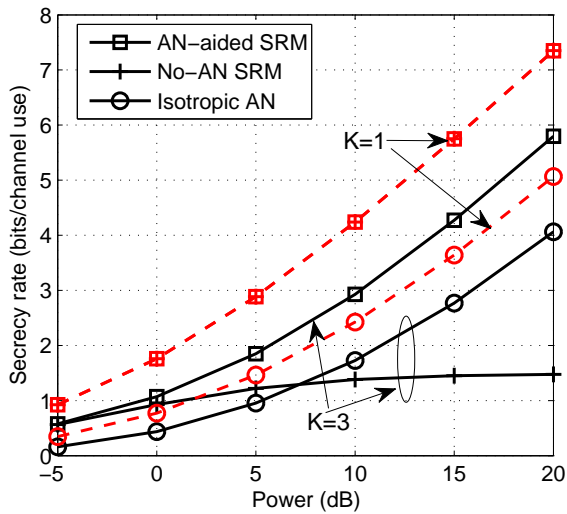}}
}
\vspace{-.5\baselineskip}
\caption{Secrecy rates versus the sum power.} \label{fig:fig2}
\vspace{-5pt}
\end{figure}

Fig.~\ref{fig:fig2} plots the secrecy rates of the various designs w.r.t. the sum power $P$.
We examine two cases, namely, one Eve ($K=1$) and three Eves ($K=3$), respectively.
The more interesting case is the three Eves case,
while the one Eve case aims to serve as a reference.
We will elaborate on this soon.
For the two cases, it can be seen that the proposed AN-aided SRM design achieves a secrecy rate performance at least no worse than the other two designs.
In fact, the performance gain of the AN-aided SRM design can be quite significant.
For example, for the case of $K=3$,
the secrecy rate gap between the AN-aided SRM design and the isotropic AN design can be close to $2$ bits per channel use when $P$ is large.
Moreover, the gap between AN-aided and no-AN SRM designs (for $K=3$) is even more dramatic.


\begin{figure}[htp!]
\begin{center}
\subfigure[][]{\resizebox{.45\textwidth}{!}{\includegraphics{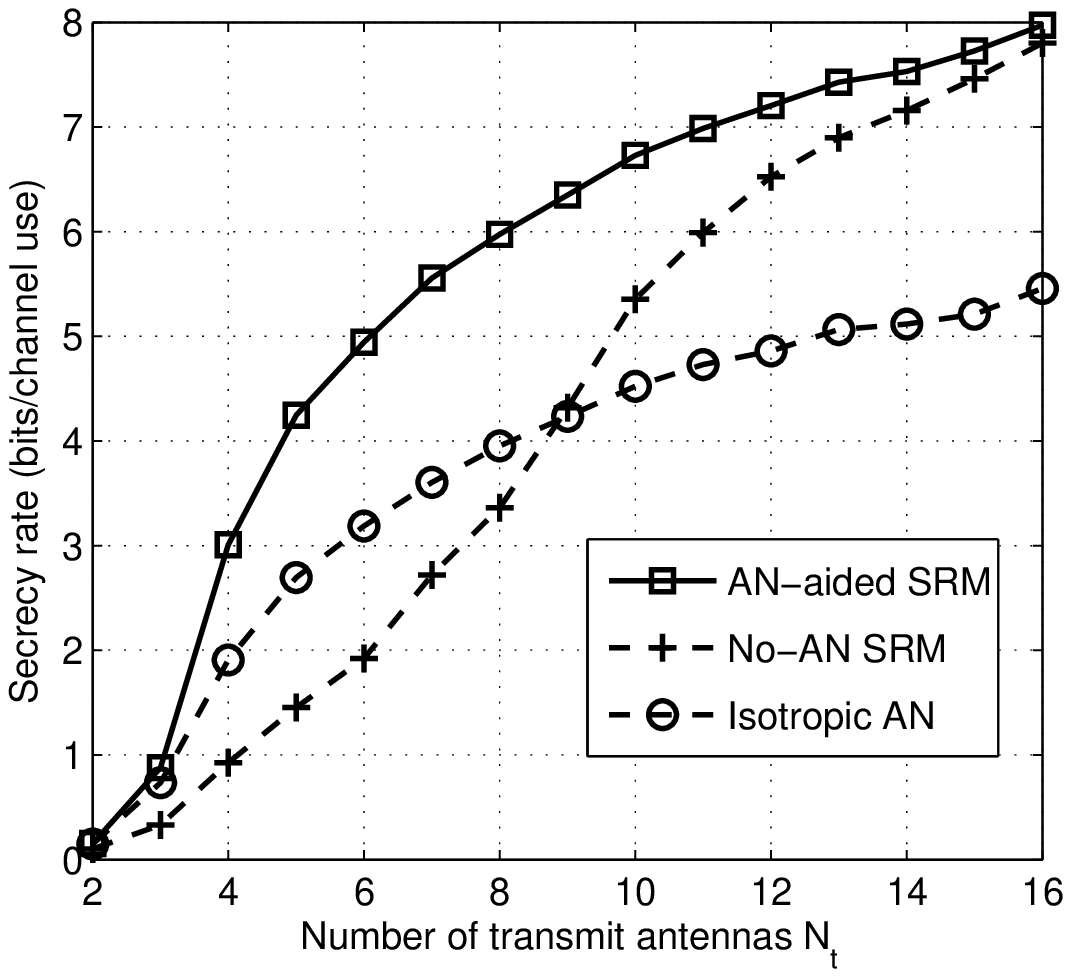}}}
\hspace{10pt}
\subfigure[][]{\resizebox{.45\textwidth}{!}{\includegraphics{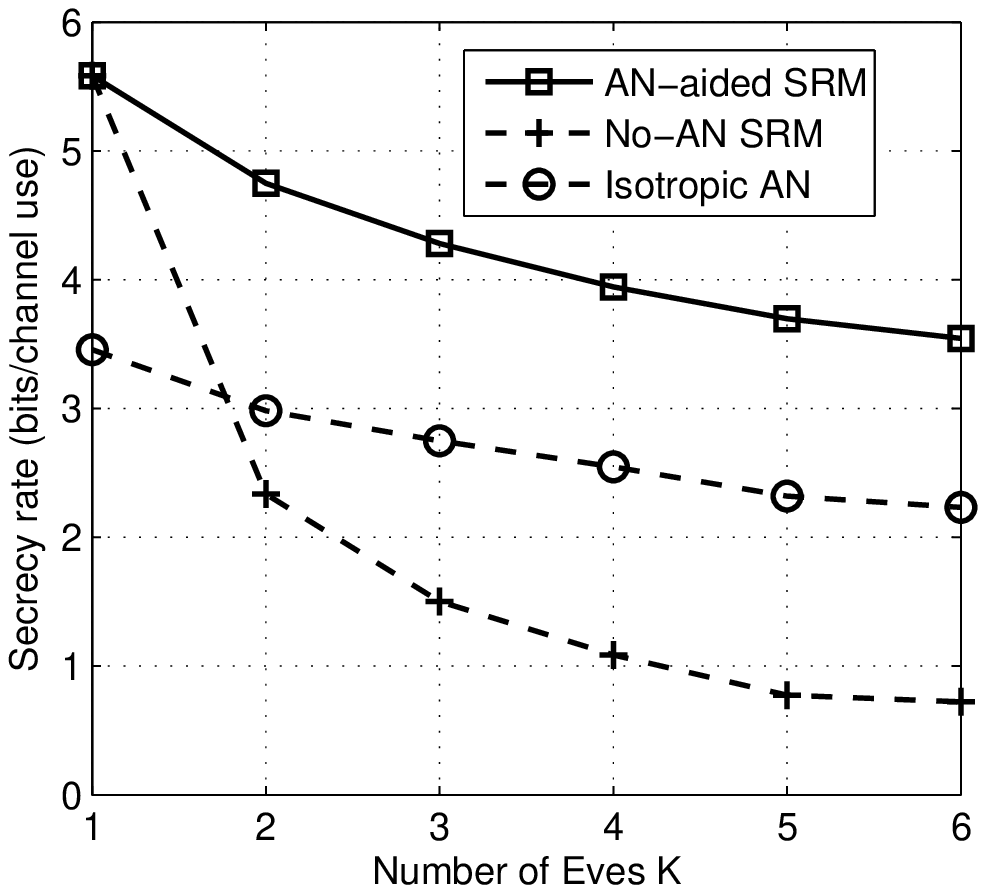}}}
\end{center}
\vspace{-.3\baselineskip}
\caption{Secrecy rates versus (a) the number of transmit antennas with $(N_{e,k}, K) = (3, 3)$, and (b) the number of Eves with $(N_t, N_{e,k}) = (5, 3)$.}
\label{fig:fig3}
\vspace{-5pt}
\end{figure}
\vspace{-.5\baselineskip}

Fig.~\ref{fig:fig2} reveals an interesting phenomenon.
We see that for $K=1$,
the secrecy rates of the AN-aided and no-AN SRM designs are exactly the same.
This means that AN may be not necessary for the one Eve case.
In fact, it is known that the no-AN SRM design achieves the secrecy capacity for the one Eve case~\cite{KW2007},
and the simulation result here has further confirmed that.
However, the picture becomes drastically different when there are more than one Eves.
For $K=3$, we can clearly observe from Fig.~\ref{fig:fig2} that the secrecy rate gap between the no-AN and AN-aided SRM designs are widening quite substantially with $P$.
In fact, the secrecy rate of the no-AN SRM design almost stays unaltered for $P \geq 10$dB, that even the isotropic AN design can provide better secrecy rate performance.
At this point, we should mention that the combined d.o.f. of Eves is $\sum_{k=1}^K N_{e,k} = 9$, which is higher than the transmit d.o.f. $N_t = 5$.
The insufficient transmit d.o.f. to deal with Eves is the reason for the unsatisfactory performance of the no-AN SRM design.
To verify this, we plot the secrecy rate versus the number of transmit antennas $N_t$ and the number of Eves $K$ in Fig.~\ref{fig:fig3}(a) and (b), respectively.
The sum power is fixed at $P= 15$dB. From Fig.~\ref{fig:fig3}(a) we can see that
the secrecy rate of the no-AN SRM design increases with $N_t$ and approaches that of AN-aided SRM design for large $N_t$;
this implies that if the transmitter has sufficiently large d.o.f., we need little or no AN.
On the other hand, from Fig.~\ref{fig:fig3}(b) we can see that
the secrecy rate of the no-AN SRM design drops rapidly with $K$.
In comparison, the AN-aided SRM design yields superior performance, even for the case of $K= 6$ (which corresponds to an Eves' combined d.o.f. of $\sum_{k=1}^K N_{e,k} = 18$). Hence, the simulation results in Fig.~\ref{fig:fig3}(a) and (b) further confirm that
incorporating AN in the transmit design is a powerful means to combat the d.o.f. bottleneck.


\subsection{Example 2: Secrecy Rate Performance with Additional Interference Temperature Constraints}

\begin{figure}[htp]
\centerline{\resizebox{.45\textwidth}{!}{\includegraphics{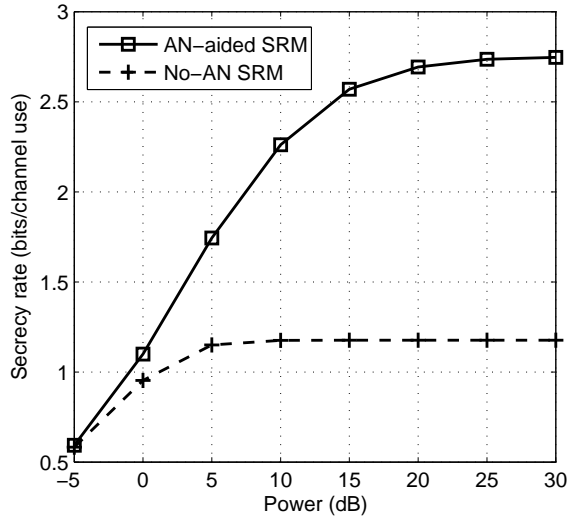}}
}
\vspace{-.5\baselineskip}
\caption{Secrecy rates versus the sum power with the interference temperature constraint $\rho = 5\,{\rm dB}$.} \label{fig:fig4}
\vspace{-20pt}
\end{figure}

This example considers SRM design with additional interference temperature constraints (ITCs).
The simulation settings are the same as the previous, namely, $N_t =5$, $N_{e,k}= 3$ for all $k$, $K= 3$, and there is one primary user with two receive antennas, i.e., $L=1$, $N_{p,1} = 2$. The primary user's channel ${\bf R}$ is generated from a standard complex Gaussian distribution.
The ITC level is set to $\rho= 5$dB; cf. Eq.~\eqref{eq:ITC}. The proposed AN-aided SRM design is benchmarked against the no-AN SRM design; see~\cite{QLI_ISPACS10}. The isotropic AN design is not applicable here since it was not designed under ITC constraints.
Fig.~\ref{fig:fig4} shows the simulation results.
We can see that the proposed AN-aided SRM design offers better secrecy rate performance, especially when $P$ is large.

\subsection{Example 3: Secrecy Rate Performance under Imperfect CSI}

In this last example, our aim is to illustrate the robustness of the WCR-SRM and OCR-SRM designs (Section~\ref{sec:partial_CSI}) when there are uncertainties with Eves' CSI.
The simulation settings are as follows:
$N_t =5$, $K=3$, $N_{e,k}=3$, sum power constraint only,
i.i.d. complex Gaussian generated ${\bf h}$ and $\bar{\bf G}_k$.

\begin{figure}[h]
\begin{center}
\subfigure[][]{\resizebox{.45\textwidth}{!}{\includegraphics{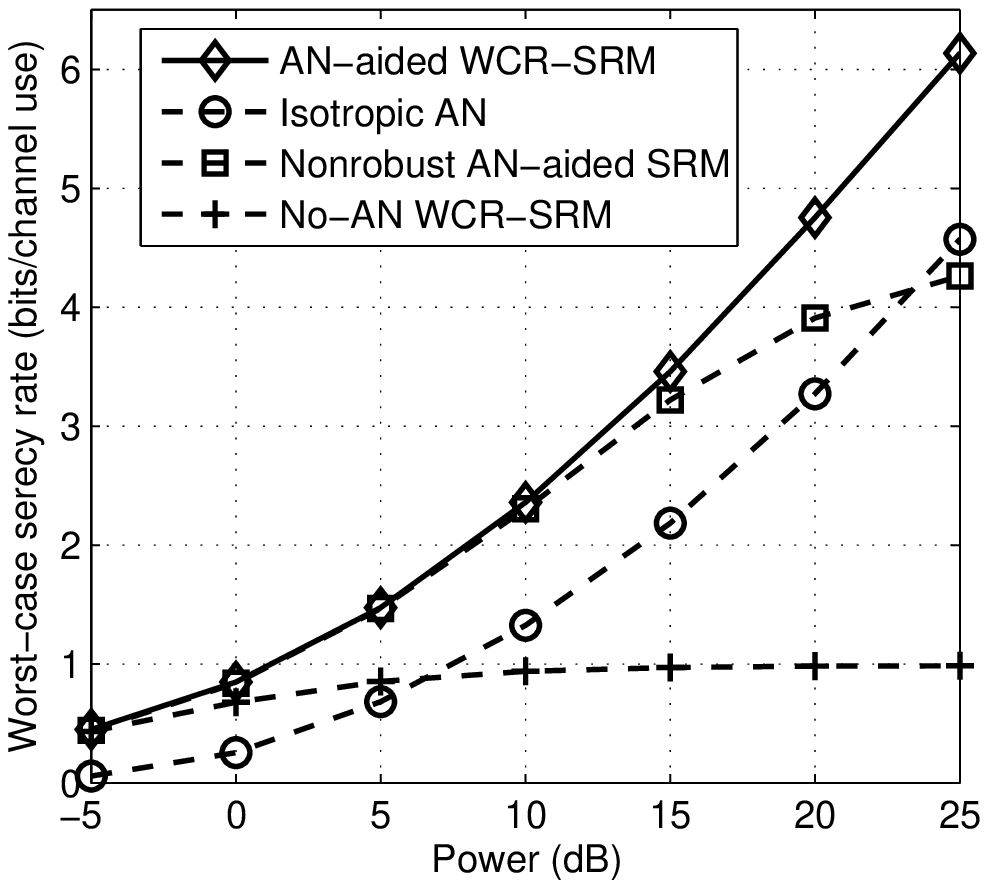}}}
\hspace{10pt}
\subfigure[][]{\resizebox{.43\textwidth}{!}{\includegraphics{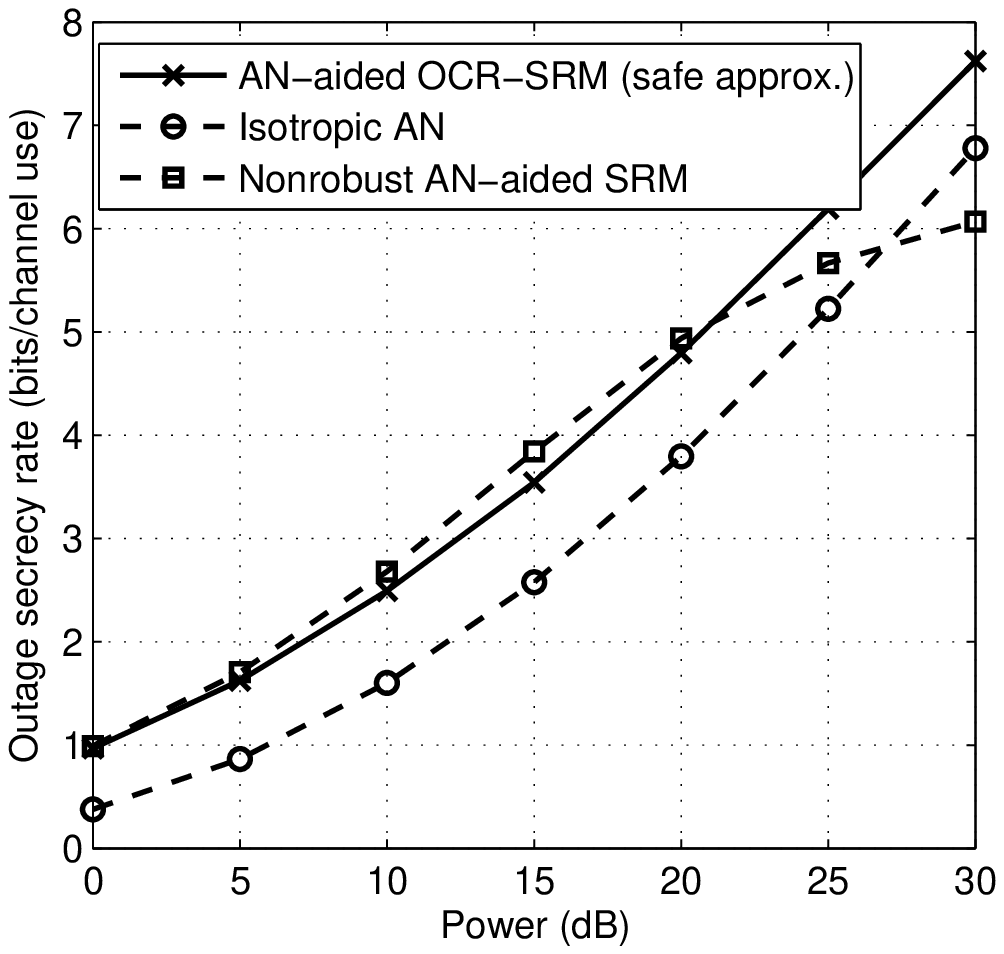}}}
\end{center}
\vspace{-.5\baselineskip}
\caption{(a) Worst-case secrecy rates versus the sum power $P$; (b) outage secrecy rates versus the sum power $P$.}
\label{fig:fig5}
\vspace{-20pt}
\end{figure}

Fig.~\ref{fig:fig5}(a) shows the performance of the various designs under the worst-case robust scenario.
We set $\varepsilon_k = 0.2$ for all $k$.
The performance measure used is the worst-case secrecy rate, which is the objective function of the WCR-SRM problem in \eqref{eq:WC_SRM_main}.
Notice that for a given design $({\bf W},\bm{\Sigma})$, the worst-case secrecy rate does not have a closed form.
In the simulation, we computed the worst-case secrecy rates of the various designs by using the WCR-SRM optimization method derived in Section~\ref{sec:wc-srm} (specifically, solving \eqref{eq:WC-SRM-relax} with $({\bf W},\bm{\Sigma})$ fixing to be a given design);
hence the development there serves a dual purpose of enabling worst-case secrecy rate computations.
In the legend of Fig.~\ref{fig:fig5}(a),
``no-AN WCR-SRM'' is the no-AN worst-case robust SRM design in \cite{QLI2011}, and ``nonrobust AN-aided SRM'' refers to the AN-aided SRM design in Section~\ref{sec:perfect_CSI}, where we apply the presumed CSIs ${\bf h}, \bar{\bf G}_1, \ldots, \bar{\bf G}_K$ (rather than the true ones) to perform the transmit design in the simulations and then evaluate the resultant worst-case secrecy rate.
From Fig.~\ref{fig:fig5}(a),
we can see that nonrobust AN-aided SRM is sensitive to CSI uncertainties.
Specifically, for $P \geq 15$dB, nonrobust AN-aided SRM exhibits performance degradation that tends to worsen as $P$ increases.
Moreover, the proposed AN-aided WCR-SRM design achieves the best worst-case secrecy rate performance compared to the other designs.

We turn our attention to the outage-constrained robust scenario.
We set $\delta =1\%$ and $\sigma_k = 0.05$ for all $k$.
The performance measure used this time is the outage-constrained secrecy rate; Monte Carlo-based evaluation was used to compute the outage-constrained secrecy rates of the considered designs.
Fig.~\ref{fig:fig5}(b) shows the outage-constrained secrecy rates of the various designs.
The results are generally consistent with their worst-case robust counterparts in Fig.~\ref{fig:fig5}(a), with one difference.
Specifically, nonrobust AN-aided SRM is seen to yield slightly better outage-constrained secrecy rate performance than AN-aided OCR-SRM for $P \leq 20$dB.
The reason is that the method for AN-aided OCR-SRM (see Section~\ref{sec:oc-srm}) is a safe approximation.
Nevertheless, nonrobust AN-aided SRM possesses the same performance degradation behavior as in the worst-case scenario,
and AN-aided OCR-SRM (by safe approximation) generally yields the best outage-constrained secrecy rate performance.

\section{Conclusion and Discussion}

This paper has considered the AN-aided secrecy rate maximization problem for an MISO channel overheard by multiple multi-antenna Eves and under both perfect and imperfect CSI.
The SRM problem is challenging to solve due to its intrinsically complex problem structures.
By resorting to an SDP-based optimization approach, we show that the SRM problem and its worst-case robust generalization can be efficiently handled by solving a sequence of SDPs.
Moreover, the development itself indicates that transmit beamforming is generally an optimal strategy for the confidential information transmission. In addition, we also propose a safe approximation to an outage-constrained robust SRM problem by
using worst-case robust SRM.
Simulation results demonstrate that the proposed designs can achieve better performance than the optimal SRM design without AN and the design with isotropic AN, especially when the number of Eves is large. These observations confirm the efficacy of AN in enhancing transmission security, as well as the necessity of optimizing AN in order to effectively interfere Eves.

%
%


As briefly mentioned in introduction, several existing AN-aided physical-layer secrecy approaches~\cite{KW2007,Negi2005,Swindlehurst2009,Mukherjee2009,XYZHOU,Jorswieck,Gerbracht12,JHuang12,JLI2012}
employ a design constraint that AN lies in the nullspace of the legitimate user's channel; i.e., ${\bf h}^H \bm\Sigma {\bf h} = 0$.
This nullspace AN constraint is reasonable, since the use of AN is intended to interfere the eavesdroppers, but not the legitimate user.
Also, using the nullspace AN constraint may help simplify design and analysis.
The AN-aided SRM problem considered here does not incorporate the nullspace AN constraint, although it is possible to do so---we can add the nullspace AN constraint to the AN-aided SRM problem (more precisely, problem~\eqref{eq:SRM_main} with an extra constraint ${\bf h}^H \bm\Sigma {\bf h} = 0$), and the resulting problem can be handled by essentially the same SDP-based optimization approach described in this paper, with some minor modification.
We found that the nullspace AN constraint does not help simplify the optimization of the SRM problem.
However, we examined by simulations that the SRM problem with the nullspace AN constraint can achieve secrecy rate performance quite close to that without the nullspace AN constraint;
the simulation results are not shown here due to the page limit.
This observation suggests that it is sound to consider the nullspace AN constraint.
As a future work, it would be interesting to analyze the SRM-optimal AN solution, studying how close it approaches the nullspace AN condition.

%

\section{Acknowledgment}
The authors would like to sincerely thank the anonymous reviewers
for their helpful and insightful comments.

\appendix

\subsection{Proof of Proposition~\ref{prop:ineq_relax_key}} \label{appendix_Prop1}
The idea of the proof is to establish some lower bounds on the left-hand side (LHS) of \eqref{eq:fact1_a}. First, note the following
equivalence
\begin{subequations}\label{appdix_equivalence}
\begin{align}
& \log \det\left( {\bf I} + \left({\bf I} + {\bf G}^H {\bm \Sigma}
{\bf G} \right)^{-1} {\bf G}^H
    {\bf W} {\bf G} \right) \leq \log \beta \label{appdix_equivalence_a} \\
  \hspace{-5pt}     \Longleftrightarrow & \det\left( {\bf I} +  {\bf U}^{-\frac{1}{2}} {\bf G}^H {\bf W} {\bf G} {\bf U}^{-\frac{1}{2}} \right) \leq
       \beta, \label{appdix_equivalence_d}
    \end{align}
\end{subequations}
where ${\bf U} = {\bf I} + {\bf G}^H {\bm \Sigma} {\bf G}$.
Eq.~\eqref{appdix_equivalence_d} is obtained by applying the basic matrix result
$\det( {\bf I} + {\bf A} {\bf B}) = \det ({\bf I} + {\bf B} {\bf
A})$
to \eqref{appdix_equivalence_a}.
Moreover, \eqref{appdix_equivalence_d} implies that
$\beta \geq 1$. To proceed further, we need the following lemma, which provides a lower
bound on the LHS of \eqref{appdix_equivalence_d}.
\begin{Lemma}[\hspace{-.5pt}\cite{QLI2011}]\label{lemma:appendix_key_ineq}
 Let ${\bf A} \succeq {\bf 0}$.
It holds true that
\begin{equation} \label{eq:trace_det_lemma}
  \det( {\bf I}+ {\bf A} ) \geq 1 + {\rm Tr}({\bf A}),
\end{equation}
and that the equality in \eqref{eq:trace_det_lemma} holds if and only if
${\rm rank}({\bf A}) \leq 1$.
\end{Lemma}
Applying Lemma~\ref{lemma:appendix_key_ineq} to the LHS of \eqref{appdix_equivalence_d} yields
\begin{equation}\label{eq:det_tr_ineq}
\det\big( {\bf I} + {\bf U}^{-\frac{1}{2}} {\bf G}^H {\bf W} {\bf G} {\bf U}^{-\frac{1}{2}} \big) \geq 1 + {\rm Tr} \big( {\bf U}^{-\frac{1}{2}} {\bf G}^H {\bf W} {\bf G} {\bf U}^{-\frac{1}{2}}  \big).
\end{equation}
Combining \eqref{appdix_equivalence} and \eqref{eq:det_tr_ineq}, we get
\begin{equation}\label{prop1_trace_relax}
  \eqref{appdix_equivalence_a} \Longrightarrow {\rm Tr} \left( {\bf U}^{-\frac{1}{2}} {\bf G}^H {\bf W} {\bf G} {\bf U}^{-\frac{1}{2}}  \right) \leq \beta - 1.
\end{equation}
In light of ${\bf U}^{-\frac{1}{2}} {\bf G}^H {\bf W} {\bf G} {\bf U}^{-\frac{1}{2}} \succeq {\bf 0}$, and the fact that ${\rm Tr} ({\bf A}) \geq \lambda_{\max} ({\bf A})$ holds for any ${\bf A} \succeq {\bf 0}$, we have
\begin{subequations} \label{Prop1_matrix_ineq}
  \begin{align}
  \eqref{appdix_equivalence_a} &~ \Longrightarrow  \lambda_{\max} \left( {\bf U}^{-\frac{1}{2}} {\bf G}^H {\bf W} {\bf G} {\bf U}^{-\frac{1}{2}} \right) \leq \beta -1,  \label{Prop1_matrix_ineq_a}  \\
    & ~ \Longleftrightarrow  {\bf U}^{-\frac{1}{2}} {\bf G}^H {\bf W} {\bf G} {\bf U}^{-\frac{1}{2}}  \preceq (\beta -1 ) {\bf I}, \label{Prop1_matrix_ineq_b} \\
    & ~ \Longleftrightarrow   (\beta -1 ) {\bf U} \succeq {\bf G}^H {\bf W} {\bf G}, \label{Prop1_matrix_ineq_c}
  \end{align}
\end{subequations}
as desired.
As for the equivalence part of Proposition~\ref{prop:ineq_relax_key}, we verify it
as follows. By Lemma~\ref{lemma:appendix_key_ineq}, the equality in \eqref{eq:det_tr_ineq} holds
if ${\rm rank}( {\bf W} ) \leq 1$. This is because ${\rm rank}( {\bf
W} ) \leq 1$ implies ${\rm rank} \big( {\bf U}^{-\frac{1}{2}} {\bf G}^H {\bf W} {\bf
G} {\bf U}^{-\frac{1}{2}} \big) \leq 1$ and thus the equality
condition in Lemma~\ref{lemma:appendix_key_ineq} holds. Next we show that
\eqref{Prop1_matrix_ineq_a} also implies the right-hand side of \eqref{prop1_trace_relax}
when ${\rm rank}({\bf W}) \leq 1$. By the following fact
\[ {\bf A} \succeq {\bf 0}, ~{\rm rank}({\bf A}) \leq 1 \Longleftrightarrow  {\bf A} = {\bm a} {\bm a}^H~ {\rm for ~some~vector }~{\bm a}, \]
we have ${\bf U}^{-\frac{1}{2}} {\bf G}^H {\bf W} {\bf G} {\bf U}^{-\frac{1}{2}} = \bm w \bm w^H$ for some vector ${\bm w} \in \mathbb{C}^{N_t}$. Therefore, \eqref{Prop1_matrix_ineq_a} can be re-expressed as $\lambda_{\max}( {\bm w} {\bm w}^H) \leq (\beta -1) $, which is
equivalent to the right-hand side of \eqref{prop1_trace_relax} by noting that $\lambda_{\max}( {\bm w} {\bm w}^H) = {\rm Tr}({\bm w} {\bm w}^H)$.


\subsection{Proof of Theorem~\ref{theorem:perfect-CSI}} \label{appendix_Thm1}
For ease of exposition, we recall the SRM problem~\eqref{eq:SRM_main1} and its relaxed problem~\eqref{eq:SRM_relax}, which are respectively given by
\begin{equation} \label{SRM_main_inner}
\begin{array}{cc}
    R^\star = \displaystyle \max_{\beta \geq 1} &  \left\{
    \begin{array}{rl}
\displaystyle \max_{{\bf W} \succeq {\bf 0}, {\bm \Sigma} \succeq {\bf 0} } ~~&  \log \left( \frac{1 + {\bf h}^H ( {\bf W} + {\bm
\Sigma} ) {\bf h}} {\beta( 1 + {\bf h}^H
{\bm \Sigma} {\bf h} ) } \right)  \\
    {\rm s.t.} ~~&  \log \det\left( {\bf I} + \left({\bf I} + {\bf G}_k^H {\bm \Sigma} {\bf G}_k \right)^{-1} {\bf G}_k^H
    {\bf W} {\bf G}_k \right) \leq \log \beta, ~ \forall k \in {\cal K} \\
    ~~&  {\rm Tr}({\bf W} + {\bm \Sigma}) \leq P, ~~{\rm Tr}({\bm \Phi}_l \left( {\bf W} + {\bm \Sigma} \right) ) \leq \rho_l,~ \forall l \in {\cal L},
    \end{array}
      \right\}
  \end{array}
\end{equation}
and
\begin{equation} \label{SRM_relax_inner}
  \begin{array}{cc}
    \bar{R}^\star = \displaystyle \max_{\beta \geq 1} &  \left\{
    \begin{array}{rl}
        \displaystyle \max_{{\bf W}\succeq {\bf 0} , {\bm \Sigma}\succeq {\bf 0}} ~~&  \log \left( \frac{1 + {\bf h}^H ( {\bf W} + {\bm
\Sigma} ) {\bf h}} {\beta( 1 + {\bf h}^H
{\bm \Sigma} {\bf h} ) } \right)  \\
    {\rm s.t.} ~~&  (\beta -1) ( {\bf I} + {\bf G}_k^H {\bm \Sigma} {\bf G}_k) - {\bf G}_k^H
    {\bf W} {\bf G}_k \succeq {\bf 0}, ~ \forall k \in {\cal K} \\
    ~~&  {\rm Tr}({\bf W} + {\bm \Sigma}) \leq P, ~~{\rm Tr}({\bm \Phi}_l \left( {\bf W} + {\bm \Sigma} \right) ) \leq \rho_l,~ \forall l\in {\cal L}
    \end{array}
      \right\}.
  \end{array}
\end{equation}
The proof consists of two steps: First, we show that for any given feasible $\beta$, there exists an optimal
${\bf W}$ for the inner maximization problem of \eqref{SRM_relax_inner} such that ${\rm rank}({\bf W}) \leq 1$;
second, we show that such a ${\bf W}$ is also optimal for the inner maximization problem of \eqref{SRM_main_inner}, and hence a solution correspondence is established between Problems~\eqref{SRM_main_inner} and \eqref{SRM_relax_inner}.

{\bf Step 1:} Given a feasible $\beta$ of \eqref{SRM_relax_inner}, let $\bar{R}_\beta$ denote the optimal value of the inner maximization problem of \eqref{SRM_relax_inner}. Consider the following power minimization problem:
\begin{subequations} \label{appendix_Pow_min}
  \begin{align}
 \min_{{\bf W} \succeq {\bf 0}, {\bm \Sigma} \succeq {\bf 0}}  &  ~ {\rm Tr}({\bf W} + {\bm \Sigma}) \label{appendix_Pow_min_a} \\
      {\rm s.t.} & ~    \log \big( \frac{1 + {\bf h}^H ( {\bf W} + {\bm \Sigma} ) {\bf h}} {\beta( 1 + {\bf h}^H
    {\bm \Sigma} {\bf h} ) } \big) \geq \bar{R}_\beta,  \label{appendix_Pow_min_b} \\
    & ~ (\beta -1 ) ({\bf I} + {\bf G}_k^H {\bm \Sigma} {\bf G}_k)  \succeq  {\bf G}_k^H {\bf W} {\bf G}_k, ~\forall k \in {\cal K}, \\
    & ~  {\rm Tr}\left({\bm \Phi}_l ({\bf W} + {\bm \Sigma}) \right) \leq \rho_l,~ \forall l \in {\cal L}.\label{appendix_Pow_min_c}
  \end{align}
\end{subequations}
Here, problem~\eqref{appendix_Pow_min} aims to minimize the total transmit power subject to a minimum requirement of the secrecy rate $\bar{R}_\beta$. The reason why we consider problem~\eqref{appendix_Pow_min} is as follows (we will prove them later): First, the optimal solution of~\eqref{appendix_Pow_min} is also optimal for the inner maximization problem of \eqref{SRM_relax_inner}; second, the optimal solution ${\bf W}$ of \eqref{appendix_Pow_min} must satisfy ${\rm rank}( {\bf W} ) \leq 1$. Combining the above two claims, the existence of an optimal ${\bf W}$ with ${\rm rank}(\bf W) \leq 1$ is readily established for the inner maximization problem of \eqref{SRM_relax_inner}.

Let $(\bar{\bf W}, \bar{\bm \Sigma})$ and $(\hat{\bf W}, \hat{\bm \Sigma})$ denote the optimal solutions of the inner maximization problems of \eqref{SRM_relax_inner} and \eqref{appendix_Pow_min}, respectively. One can easily verify that $(\bar{\bf W}, \bar{\bm \Sigma})$ is a feasible solution of \eqref{appendix_Pow_min}. It follows that
\begin{equation}\label{appendix_Thm1_ineq_1}
  {\rm Tr}(\hat{\bf W} + \hat{\bm \Sigma}) \leq  {\rm Tr}(\bar{\bf W} + \bar{\bm \Sigma})\leq P,
\end{equation}
where the first inequality is due to the fact that $(\hat{\bf W}, \hat{\bm \Sigma})$ minimizes ${\rm Tr}(\hat{\bf W} + \hat{\bm \Sigma})$ (cf. Problem \eqref{appendix_Pow_min_a}); the second inequality follows from the feasibility of $(\bar{\bf W}, \bar{\bm \Sigma})$ w.r.t.
\eqref{SRM_relax_inner}. The inequality~\eqref{appendix_Thm1_ineq_1}, together with
\eqref{appendix_Pow_min_c}, imply that $(\hat{\bf W}, \hat{\bm
\Sigma})$ is a feasible solution of
\eqref{SRM_relax_inner}, i.e.,
\begin{equation}\label{appendix_Thm1_ineq_2}
\log \left( \frac{ 1+  {\bf h}^H (\hat{\bf W} + \hat{\bm \Sigma}) {\bf
h}}{\beta ( 1 + {\bf h}^H \hat{\bm \Sigma} {\bf h}
    )}  \right)  \leq  \bar{R}_{\beta}.
\end{equation}
Combining \eqref{appendix_Thm1_ineq_2} with \eqref{appendix_Pow_min_b} yields
\[ \log \left( \frac{ 1+  {\bf h}^H (\hat{\bf W} + \hat{\bm \Sigma}) {\bf
h}}{\beta ( 1 + {\bf h}^H  \hat{\bm \Sigma} {\bf h}
    )}  \right)  =  \bar{R}_{\beta}. \]
Therefore, $(\hat{\bf W}, \hat{\bm \Sigma})$ is an optimal
solution of the inner maximization problem of \eqref{SRM_relax_inner}.

To show ${\rm rank} (\hat{\bf W}) \leq 1$, we check the Karush-Kuhn-Tucker (KKT) optimality conditions of problem~\eqref{appendix_Pow_min}. Let us rewrite
\eqref{appendix_Pow_min} as
\begin{subequations} \label{eq:appdix_PM_eqv}
  \begin{align}
    \min_{{\bf W} \succeq {\bf 0}, {\bm \Sigma} \succeq {\bf 0}} & ~ {\rm Tr}({\bf W} + {\bm \Sigma}) \label{eq:appdix_PM_a} \\
    {\rm s.t.} & ~  {\bf h}^H \left ( {\bf W} + \mu {\bm \Sigma}
    \right) {\bf h}
    + \mu \geq 0, \label{eq:appdix_PM_eqv_b} \\
    & ~ {\bf T}_k({\bf W}, {\bm \Sigma}) \succeq {\bf 0}, ~ {\rm Tr}\left( {\bm \Phi}_l ( {\bf W} + {\bm \Sigma}) \right) \leq \rho_l,
~ \forall k, l,  \label{eq:appdix_PM_eqv_c}
  \end{align}
\end{subequations}
where $\mu = 1- \beta 2^{\bar{R}_\beta} $, and $ {\bf T}_k({\bf W}, {\bm
\Sigma}) \triangleq (\beta -1 ) ({\bf I} + {\bf G}_k^H {\bm \Sigma} {\bf G}_k
) - {\bf G}_k^H {\bf W} {\bf G}_k, ~\forall k.$
The Lagrangian of problem~\eqref{eq:appdix_PM_eqv} is given by
\begin{align*}
   \mathcal{L} ( \mathcal{X} ) =  & {\rm Tr}({\bf W} + {\bm \Sigma})  + \textstyle \sum_{l=1}^L \eta_l \big( {\rm Tr}\left( {\bm \Phi}_l ( {\bf W} + {\bm \Sigma}) \right)- \rho_l \big) \\
  &  - \lambda \big( {\bf h}^H ({\bf W} + \mu {\bm \Sigma})
 {\bf h} + \mu  \big) - \textstyle \sum_{k=1}^{K} {\rm Tr}\big( {\bf A}_k {\bf T}_k({\bf W}, {\bm
\Sigma})\big)  - {\rm Tr}({\bf Q} {\bf W}) - {\rm Tr}({\bf
M} {\bm \Sigma}),
\end{align*}
where $\mathcal{X}$ denotes a collection of all the primal and dual variables of problem~\eqref{eq:appdix_PM_eqv}; ${\bf Q} \in \mathbb{H}^{N_t}_+$, ${\bf M} \in \mathbb{H}^{N_t}_+ $,
$\lambda \in \mathbb{R}_+ $, ${\bf A}_k \in \mathbb{H}^{N_{e,k}}_+$ and $\eta_l \in \mathbb{R}_+$ are dual variables associated with ${\bf W} \succeq {\bf 0}$, ${\bm \Sigma} \succeq {\bf 0}$, \eqref{eq:appdix_PM_eqv_b}, and \eqref{eq:appdix_PM_eqv_c}, respectively. Assuming that problem~\eqref{eq:appdix_PM_eqv} satisfies some constraint qualifications~\cite{Bertsekas}, the KKT conditions that are
relevant to the proof are given by
\begin{subequations}\label{eq:appdix_PM_kkt}
  \begin{align}
    & \hspace{-.2cm} {\bf I} - \lambda {\bf h} {\bf h}^H +  \textstyle \sum_{k=1}^{K} {\bf G}_k
    {\bf A}_k {\bf G}_k^H  +  \textstyle \sum_{l=1}^{L} \eta_l {\bm \Phi}_l - {\bf Q}  = {\bf 0}, \label{eq:appdix_PM_kkt_a}\\
    &  {\bf Q} {\bf W}  = {\bf 0},  \label{eq:appdix_PM_kkt_b} \\
    & {\bf W} \succeq {\bf 0}, \quad {\bf A}_k \succeq {\bf 0},~\forall k, \quad \eta_l  \geq 0 , \forall l, \label{eq:appdix_PM_kkt_c}
  \end{align}
\end{subequations}
Postmultiplying \eqref{eq:appdix_PM_kkt_a} by ${\bf W}$ and making
use of \eqref{eq:appdix_PM_kkt_b} yield
\begin{equation}\label{eq:appdix_PM_kkt_multiply}
 \big( {\bf I} + \textstyle \sum_{k=1}^{K} {\bf G}_k
    {\bf A}_k {\bf G}_k^H + \textstyle \sum_{l=1}^{L} \eta_l {\bm \Phi}_l \big) {\bf W} = \lambda {\bf h} {\bf
    h}^H  {\bf W},
\end{equation}
which implies that
\begin{equation}\label{eq:appdix1_rank_relation_a}
\begin{aligned}
& {\rm rank} \big( ({\bf I} + \textstyle \sum_{k=1}^{K} {\bf G}_k
    {\bf A}_k {\bf G}_k^H  +  \textstyle \sum_{l=1}^{L} \eta_l {\bm \Phi}_l ) {\bf W} \big) = {\rm rank}(\lambda {\bf h} {\bf h}^H {\bf W} ) \leq 1.
\end{aligned}
\end{equation}
Since ${\bf I} +
\sum_{k=1}^{K} {\bf G}_k {\bf A}_k {\bf G}_k^H  + \sum_{l=1}^{L} \eta_l {\bm \Phi}_l \succ {\bf 0}$, the
following relation holds
\begin{equation}\label{eq:appdix1_rank_relation_b}
  {\rm rank}({\bf W}) = {\rm rank}\big( ({\bf I} + \textstyle\sum_{k=1}^{K} {\bf G}_k
    {\bf A}_k {\bf G}_k^H  + \sum_{l=1}^{L} \eta_l {\bm \Phi}_l )  {\bf W} \big).
\end{equation}
Finally, using \eqref{eq:appdix1_rank_relation_a} and
\eqref{eq:appdix1_rank_relation_b} produces the desired result ${\rm rank}({\bf W}) \leq 1$.

{\bf Step 2:} Let $\phi_\beta({\bf W}, {\bm \Sigma})$ be the objective function of the
inner maximization problem of \eqref{SRM_main_inner} (or problem \eqref{SRM_relax_inner}) for a particular $\beta$, and $(\breve{\bf W}, \breve{\bm \Sigma})$ and $(\bar{\bf W}, \bar{\bm \Sigma})$ be the corresponding optimal solutions of the inner maximization problem of \eqref{SRM_main_inner} and \eqref{SRM_relax_inner}, respectively. Without loss of generality, we assume ${\rm rank}(\bar{\bf W}) \leq 1$.
Since the inner maximization problem of
\eqref{SRM_relax_inner} is a relaxation of that of
\eqref{SRM_main_inner} (cf. Proposition~\ref{prop:ineq_relax_key}), we have
\[
  \phi_\beta(\bar{\bf W}, \bar{\bm \Sigma}) \geq  \phi_\beta(\breve{\bf W}, \breve{\bm
  \Sigma}).
\]
On the other hand, the condition ${\rm rank}(\bar{\bf W}) \leq 1$
implies that $(\bar{\bf W}, \bar{\bm \Sigma})$ is also a feasible
solution of the inner maximization problem of \eqref{SRM_main_inner}, owing to the equivalence condition in Proposition~\ref{prop:ineq_relax_key}. As a result, we have
\[   \phi_\beta(\bar{\bf W}, \bar{\bm \Sigma}) \leq  \phi_\beta(\breve{\bf W}, \breve{\bm
  \Sigma}). \]
Combining the above two inequalities, we conclude that
$\phi_\beta(\bar{\bf W}, \bar{\bm \Sigma}) = \phi_\beta(\breve{\bf W}, \breve{\bm
  \Sigma})$, i.e., $(\bar{\bf W}, \bar{\bm \Sigma})$ is also optimal for the inner maximization problem of \eqref{SRM_main_inner}.

We have established a solution correspondence between \eqref{SRM_main_inner} and \eqref{SRM_relax_inner} for any given feasible $\beta$, which includes the optimal $\beta^\star$. Subsequently, the results in Theorem~\ref{theorem:perfect-CSI} are obtained.

\subsection{Proof of Theorem~\ref{theorem_WC}} \label{appendix:theorem-wc}
The proof is reminiscent of Theorem~\ref{theorem:perfect-CSI}. Re-express \eqref{eq:WC-SRM-relax} as


\begin{equation}  \label{appendix-WC-relax-eqv}
\begin{array}{cc}
   \bar{R}^\star = \displaystyle \max_{\beta \geq 1} &  \left\{
    \begin{array}{rl}
 &  \displaystyle \max_{{\bf W}, {\bm \Sigma}, \{t_k\}} ~  \log \left(\frac{1 + {\bf h}^H ( {\bf W} + {\bm \Sigma} ) {\bf h}}{\beta ( 1 + {\bf h}^H {\bm \Sigma} {\bf h} ) } \right)  \\
& \hspace{0.5cm} {\rm s.t.}  ~ {\bf T}_k (\beta, {\bf W}, {\bm \Sigma}, t_k ) \succeq {\bf 0},\,t_k \geq 0, \,\forall k \in {\cal K}, \\
            &  \hspace{1.2cm}  {\rm Tr}\left( {\bm \Phi}_l ( {\bf W} + {\bm \Sigma} ) \right) \leq \rho_l,~ \forall l \in {\cal L}, \\
            &  \hspace{1.2cm} {\rm Tr}({\bf W} + {\bm \Sigma}) \leq P,~{\bf W}\succeq {\bf 0},~ {\bm \Sigma}\succeq {\bf 0}
\end{array}
\right \}
\end{array}
\end{equation}
Given a feasible $\beta$ of \eqref{appendix-WC-relax-eqv}, let $\bar{R}_\beta$ denote the optimal value of the inner maximization problem of \eqref{appendix-WC-relax-eqv}. Consider the following secrecy-rate constrained power minimization problem
\begin{equation} \label{appendix-wc-power-min}
  \begin{aligned}
   \min_{{\bf W}\succeq {\bf 0}, {\bm \Sigma}\succeq {\bf 0}, t_1,\ldots, t_K} & ~ {\rm Tr}({\bf W} + {\bm \Sigma}) \\
    {\rm s.t.} & ~  \log \big( \frac{ 1+  {\bf h}^H ({\bf W + {\bm \Sigma}}) {\bf h}} { \beta ( 1 + {\bf h}^H {\bm \Sigma} {\bf h}
    ) }  \big) \geq \bar{R}_\beta, \\
    &  ~ {\bf T}_k(\beta, {\bf W}, {\bm \Sigma}, t_k ) \succeq {\bf 0},  \quad t_k \geq 0, ~\forall k \in {\cal K}, \\
    &  ~ {\rm Tr}\left({\bm \Phi}_l ({\bf W} + {\bm \Sigma} ) \right) \leq \rho_l,~ \forall l\in {\cal L}.
  \end{aligned}
\end{equation}
Following the same argument in the proof of Theorem~\ref{theorem:perfect-CSI}, one can easily verify that the optimal solution of \eqref{appendix-wc-power-min} must be optimal for the inner maximization problem of
\eqref{appendix-WC-relax-eqv}. Next, we show that the optimal solution of \eqref{appendix-wc-power-min} has a rank no greater than one by checking its KKT conditions. Rewrite \eqref{appendix-wc-power-min} as
\begin{subequations} \label{eq:WC_pow_LMI_eqv}
  \begin{align}
\min_{{\bf W} , {\bm \Sigma} , \{ t_k\}_{k\in{\cal K}}} & ~ {\rm Tr}({\bf W} + {\bm \Sigma})  \label{eq:WC_pow_LMI_eqv_a} \\
    {\rm s.t.} & ~  {\bf h }^H ( {\bf W} + \mu {\bm \Sigma} ) {\bf h} + \mu \geq 0, \label{eq:WC_pow_LMI_eqv_b}\\
    & ~ \tilde{\bf G}_k^H \left( (\beta -1) {\bm \Sigma}
    - {\bf W} \right) \tilde{\bf G}_k + \begin{bmatrix}
(\beta - 1- t_k) {\bf I} &  {\bf 0} \\
{\bf 0} & \frac{t_k}{\epsilon_k^2} {\bf I}
    \end{bmatrix}  \succeq {\bf 0}, ~ t_k \geq 0, ~\forall k \in {\cal K} \label{eq:WC_pow_LMI_eqv_c} \\
    & ~ {\rm Tr} \left({\bm \Phi}_l ({\bf W} + {\bm \Sigma}) \right) \leq \rho_l,~\forall l \in {\cal L}, \label{eq:WC_pow_LMI_eqv_d} \\
    & ~ {\bf W}\succeq {\bf 0},~~ {\bm \Sigma}\succeq {\bf 0}, \label{eq:WC_pow_LMI_eqv_e}
  \end{align}
\end{subequations}
where $\mu = 1- 2^{\bar{R}_\beta} \beta$ and $\tilde{\bf G}_k = [~ \bar{\bf G}_k,~
{\bf I}~]$. The Lagrangian of problem~\eqref{eq:WC_pow_LMI_eqv} is given by
\setlength{\arraycolsep}{2pt}
\[\begin{array}{rl}
 \mathcal{L} (\mathcal{X}) = & {\rm Tr}( {\bf W} + {\bm \Sigma} ) - \sum_{k=1}^{K} {\rm Tr}({\bf A}_k {\bf T}_k (\beta, {\bf W}, {\bm \Sigma}, t_k)) \\
  &\hspace{-10pt} + \sum_{l=1}^{L} \eta_l \left( {\rm Tr} \left({\bm \Phi}_l ({\bf W} + {\bm \Sigma}) \right) - \rho_l \right) - {\rm Tr}({\bm \Sigma} {\bf M}) \\
 &\hspace{-10pt}  - {\rm Tr}({\bf W} {\bf Q}) - \nu \left( {\bf h}^H ({\bf W} + \mu {\bm \Sigma}) {\bf h} + \mu \right)   - \sum_{k=1}^{K} \lambda_k t_k,
 \end{array} \]
 where $\mathcal{X}$ denotes a collection of all primal and dual variables; ${\bf Q} \in \mathbb{H}^{N_t}_+$, ${\bf M} \in \mathbb{H}^{N_t}_+$, $\nu \in \mathbb{R}_+$, ${\bf A}_k \in \mathbb{H}^{N_{e,k}+N_t}_+$, $\lambda_k \in \mathbb{R}_+, \forall k$ and $\eta_l \in \mathbb{R}_+, \forall l $ are dual variables associated
with ${\bf W}$, ${\bm \Sigma}$, \eqref{eq:WC_pow_LMI_eqv_b}, \eqref{eq:WC_pow_LMI_eqv_c} and \eqref{eq:WC_pow_LMI_eqv_d}, respectively.
Parts of the KKT conditions of problem \eqref{eq:WC_pow_LMI_eqv} are listed
below:
\begin{subequations}\label{eq:theorem3_kkt}
  \begin{align}
    & {\bf I} + \sum_{k=1}^{K} \tilde{\bf G}_k {\bf A}_k \tilde{\bf
    G}_k^H + \sum_{l=1}^{L} \eta_l {\bm \Phi}_l - \nu {\bf h} {\bf h}^H  - {\bf Q}  = {\bf 0},     \label{eq:theorem3_kkt_a}\\
& {\bf Q} {\bf W}    = {\bf 0}, \label{eq:theorem3_kkt_b}\\
& {\bf W} \succeq {\bf 0}, \quad {\bf Q} \succeq {\bf 0}, \quad {\bf A}_k \succeq {\bf 0},\forall k \in {\cal K}, \quad \eta_l \geq 0, \forall l \in {\cal L},  \label{eq:theorem3_kkt_c}
  \end{align}
\end{subequations}
Postmultiplying \eqref{eq:theorem3_kkt_a} by ${\bf W}$ and making
use of \eqref{eq:theorem3_kkt_b}, we obtain
\[
  ({\bf I} + \textstyle \sum_{k=1}^{K} \tilde{\bf G}_k {\bf A}_k \tilde{\bf G}_k^H
  + \sum_{l=1}^{L} \eta_l {\bm \Phi}_l) {\bf W} = \nu {\bf h} {\bf h}^H {\bf W},
\]
which implies that
\begin{equation}\label{eq:theorem3_rank_equality}
\begin{aligned}
  & {\rm rank} \big(  \big({\bf I} + \textstyle \sum_{k=1}^{K} \tilde{\bf
G}_k {\bf A}_k \tilde{\bf G}_k^H  + \sum_{l=1}^{L} \eta_l {\bm \Phi}_l \big) {\bf W} \big) = {\rm rank}(\nu {\bf h} {\bf h}^H {\bf W}) \leq 1.
\end{aligned}
\end{equation}
Since ${\bf I} + \textstyle \sum_{k=1}^{K} \tilde{\bf G}_k {\bf A}_k
\tilde{\bf G}_k^H + \sum_{l=1}^{L} \eta_l {\bm \Phi}_l \succ {\bf 0} $, it holds that
\begin{equation} \label{appendix:theorem-wc-rank-equ}
\begin{aligned}
  & {\rm rank}({\bf W}) =  {\rm rank} \big(  ({\bf I} +
\textstyle \sum_{k=1}^{K} \tilde{\bf G}_k {\bf A}_k \tilde{\bf
G}_k^H  + \sum_{l=1}^{L} \eta_l {\bm \Phi}_l) {\bf W} \big).
\end{aligned}
\end{equation}
Combining \eqref{appendix:theorem-wc-rank-equ} and \eqref{eq:theorem3_rank_equality} yields ${\rm rank}({\bf W}) \leq 1$.

To complete the proof, we still need to argue that such an optimal ${\bf W}$ is also optimal for \eqref{eq:WC-SRM-reform} for the same $\beta$. The proof is essentially identical to that of Theorem~\ref{theorem:perfect-CSI}, and thus is omitted for brevity.


\begin{thebibliography}{10}
\providecommand{\url}[1]{#1}
\csname url@samestyle\endcsname
\providecommand{\newblock}{\relax}
\providecommand{\bibinfo}[2]{#2}
\providecommand{\BIBentrySTDinterwordspacing}{\spaceskip=0pt\relax}
\providecommand{\BIBentryALTinterwordstretchfactor}{4}
\providecommand{\BIBentryALTinterwordspacing}{\spaceskip=\fontdimen2\font plus
\BIBentryALTinterwordstretchfactor\fontdimen3\font minus
  \fontdimen4\font\relax}
\providecommand{\BIBforeignlanguage}[2]{{%
\expandafter\ifx\csname l@#1\endcsname\relax
\typeout{** WARNING: IEEEtran.bst: No hyphenation pattern has been}%
\typeout{** loaded for the language `#1'. Using the pattern for}%
\typeout{** the default language instead.}%
\else
\language=\csname l@#1\endcsname
\fi
#2}}
\providecommand{\BIBdecl}{\relax}
\BIBdecl

\bibitem{Wyner1975}
A.~D. Wyner, ``The wiretap channel,'' in \emph{The Bell System Technical
  Journal}, vol.~54, October 1975, pp. 1355--1387.

\bibitem{Liang_book}
Y.~Liang, H.~V. Poor, and S.~Shamai{ }(Shitz), ``Information theoretic
  security,'' \emph{Foundations and Trends in Communications and Information
  Theory}, vol.~5, no. 4-5, pp. 355--580, 2008.

\bibitem{Negi2005}
R.~Negi and S.~Goel, ``Secret communication using artificial noise,'' in
  \emph{IEEE Vehicular Technology Conference (VTC)}, Sept. 2005, pp.
  1906--1910.

\bibitem{KW2007}
A.~Khisti and G.~W. Wornell, ``Secure transmission with multiple antennas {I}:
  The {MISOME} wiretap channel,'' \emph{IEEE Trans. Inform. Theory}, vol.~56,
  no.~7, pp. 3088--3104, July 2010.

\bibitem{Swindlehurst2009}
A.~L. Swindlehurst, ``Fixed {SINR} solution for the {MIMO} wiretap channel,''
  in \emph{Proc. IEEE Int. Conf. Acoustics, Speech, and Signal Processing
  (ICASSP) 2009}, April 2009, pp. 2437--2440.

\bibitem{Mukherjee2009}
A.~Mukherjee and A.~L. Swindlehurst, ``Fixed-rate power allocation strategies
  for enhanced secrecy in {MIMO} wiretap channels,'' in \emph{Proc. IEEE Signal
  Process. Advances in Wireless Commun. (SPAWC) 2009}, June 2009, pp. 344--348.

\bibitem{XYZHOU}
X.~Zhou and M.~R. McKay, ``Secure transmission with artificial noise over
  fading channels: Achievable rate and optimal power allocation,'' \emph{IEEE
  Trans. Veh. Tech.}, vol.~59, no.~8, pp. 3831--3842, Oct. 2010.

\bibitem{Jorswieck}
S.~Gerbracht, A.~Wolf, and E.~A. Jorswieck, ``Beamforming for fading wiretap
  channels with partial channel information,'' in \emph{International ITG
  Workshop on Smart Antennas, Bremen, Germany}, Feb. 2010.

\bibitem{Jorswieck10}
E.~A. Jorswieck, ``Secrecy capacity of single- and multi-antenna channels with
  simple helpers,'' in \emph{Proc. of International ITG Conference on Source
  and Channel Coding (SCC)}, Jan. 2010.

\bibitem{Mukherjee12}
A.~Mukherjee and A.~L. Swindlehurst, ``Detecting passive eavesdroppers in the
  {MIMO} wiretap channel,'' in \emph{Proc. IEEE Int. Conf. Acoustics, Speech,
  and Signal Processing (ICASSP) 2012, Kyoto, Japan}, Mar. 2012.

\bibitem{Ali2011}
S.~Fakoorian and A.~L. Swindlehurst, ``Solution for the {MIMO} {G}aussian
  wiretap channel with a cooperative jammer,'' \emph{IEEE Trans. Signal
  Process.}, vol.~59, no.~10, pp. 5013--5022, Oct. 2011.

\bibitem{Liao10}
W.-C. Liao, T.-H. Chang, W.-K. Ma, and C.-Y. Chi, ``Qo{S}-based transmit
  beamforming in the presence of eavesdroppers: {A}n optimized
  artificial-noise-aided approach,'' \emph{IEEE Trans. Signal Process.},
  vol.~59, no.~3, pp. 1202--1216, Mar. 2011.

\bibitem{JHuang12}
J.~Huang and A.~L. Swindlehurst, ``Robust secure transmission in {MISO}
  channels based on worst-case optimization,'' \emph{IEEE Trans. Signal
  Process.}, vol.~60, no.~4, pp. 1696--1707, Apr. 2012.

\bibitem{Wolf10}
A.~Wolf and E.~A. Jorswieck, ``Maximization of worst-case secrecy rates in
  {MIMO} wiretap channels,'' in \emph{Proc. of the Asilomar Conference on
  Signals, Systems, and Computers, Pacific Grove, USA}, Nov. 2010.

\bibitem{QLI2011}
Q.~Li and W.-K. Ma, ``Optimal and robust transmit designs for {MISO} channel
  secrecy by semidefinite programming,'' \emph{IEEE Trans. Signal Process.},
  vol.~59, no.~8, pp. 3799--3812, Aug. 2011.

\bibitem{Gerbracht12}
S.~Gerbracht, C.~Scheunert, and E.~A. Jorswieck, ``Secrecy outage in {MISO}
  systems with partial channel information,'' \emph{IEEE Trans. Information
  Forensics and Security}, vol.~7, no.~2, pp. 704--716, April 2012.

\bibitem{JLI2012}
S.~Luo, J.~Li, and A.~P. Petropulu, ``Outage constrained secrecy rate
  maximization using cooperative jamming,'' in \emph{IEEE Statistical Signal
  Processing Workshop (SSP)}, Aug. 2012.

\bibitem{Jli2011}
J.~Li, A.~P. Petropulu, and S.~Weber, ``On cooperative relaying schemes for
  wireless physical layer security,'' \emph{IEEE Trans. Signal Process.},
  vol.~59, no.~10, pp. 4985--4997, Oct. 2011.

\bibitem{GanZheng11}
G.~Zheng, L.-C. Choo, and K.-K. Wong, ``Optimal cooperative jamming to enhance
  physical layer security using relays,'' \emph{IEEE Trans. Signal Process.},
  vol.~59, no.~3, pp. 1317--1322, Mar. 2011.

\bibitem{yiyangpei2010}
Y.~Pei, Y.-C. Liang, L.~Zhang, K.~Teh, and K.~H. Li, ``Secure communication
  over {MISO} cognitive radio channels,'' \emph{IEEE Trans. Wireless Commun.},
  vol.~9, no.~4, pp. 1494--1502, Apr. 2010.

\bibitem{Pei2011}
Y.~Pei, Y.-C. Liang, K.~Teh, and K.~H. Li, ``Secure communication in
  multiantenna cognitive radio networks with imperfect channel state
  information,'' \emph{IEEE Trans. Signal Process.}, vol.~59, no.~4, pp.
  1683--1693, Apr. 2011.

\bibitem{Gursoy2010}
J.~Zhang and M.~C. Gursoy, ``Collaborative relay beamforming for secrecy,'' in
  \emph{Proc. IEEE Int. Conf. Communications (ICC)}, 2010.

\bibitem{MinyanPei12}
M.~Pei, J.~Wei, K.-K. Wong, and X.~Wang, ``Masked beamforming for multiuser
  {MIMO} wiretap channels with imperfect {CSI},'' \emph{IEEE Trans. Wireless
  Commun.}, vol.~11, no.~2, pp. 544--549, Feb. 2012.

\bibitem{Sturm1999}
J.~Sturm, ``Using {S}e{D}u{M}i 1.02, a {MATLAB} toolbox for optimization over
  symmetric cones,'' \emph{Optim. Methods Softw.}, vol.~11, pp. 625--653, 1999,
  (webpage and software) http://sedumi.ie.lehigh.edu/.

\bibitem{Grant2011}
M.~Grant and S.~Boyd, ``{CVX}: {M}atlab software for disciplined convex
  programming,'' Apr. 2011, available online at {\urlstyle{tt}
  \url{http://cvxr.com/cvx/}}.

\bibitem{Shafiee2007}
S.~Shafiee and S.~Ulukus, ``Achievable rates in {G}aussian {MISO} channels with
  secrecy constraints,'' in \emph{IEEE Int'l Symp. on Inform. Theory}, June
  2007, pp. 2466--2470.

\bibitem{Liang2007}
Y.~Liang, G.~Kramer, H.~V. Poor, and S.~Shamai{ }(Shitz), ``Compound wire-tap
  channels,'' in \emph{Proc. 45th Annual Allerton Conf. Commun., Control, and
  Computing}, Sept. 2007, pp. 136--143.

\bibitem{WEIYU}
W.~Yu and T.~Lan, ``Transmitter optimization for the multi-antenna downlink
  with per-antenna power constraints,'' \emph{IEEE Trans. Signal Process.},
  vol.~55, no.~6, pp. 2646--2660, June 2007.

\bibitem{Huh10}
H.~Huh, H.~C. Papadopoulos, and G.~Caire, ``Multiuser {MISO} transmitter
  optimization for intercell interference mitigation,'' \emph{IEEE Trans.
  Signal Process.}, vol.~58, no.~8, pp. 4272--4285, Aug. 2010.

\bibitem{Charnes1962}
A.~Charnes and W.~W. Cooper, ``Programming with linear fractional
  functionals,'' \emph{Naval Res. Logistics Quarterly}, vol.~9, pp. 181--186,
  1962.

\bibitem{Kolda03}
T.~Kolda, R.~Lewis, and V.~Torczon, ``Optimization by direct search: new
  perspectives on some classical and modern methods,'' \emph{SIAM Review},
  vol.~45, no.~3, pp. 385--482, 2003.

\bibitem{Conn-book09}
A.~R. Conn, K.~Scheinberg, and L.~N. Vicente, \emph{Introduction to
  derivative-free optimization}.\hskip 1em plus 0.5em minus 0.4em\relax
  Philadelphia: MPS-SIAM Series on Optimization, 2009.

\bibitem{Bertsekas}
D.~Bertsekas, \emph{Nonlinear Programming}.\hskip 1em plus 0.5em minus
  0.4em\relax Belmont, MA: Athena Scientific, 1999.

\bibitem{luo04}
Z.-Q. Luo, J.~F. Sturm, and S.~Zhang, ``Multivariate nonnegative quadratic
  mappings,'' \emph{SIAM J. Optim.}, vol.~14, no.~4, pp. 1140--1162, 2004.

\bibitem{Bloch08}
M.~Bloch, J.~Barros, M.~R.~S. Rodrigues, and S.~W. McLaughlin, ``Wireless
  information-theoretic security,'' \emph{IEEE Trans. Inform. Theory}, vol.~54,
  no.~6, pp. 2515--2534, June 2008.

\bibitem{KYWang11}
K.-Y. Wang, A.~M.-C. So, T.-H. Chang, W.-K. Ma, and C.-Y. Chi, ``Outage
  constrained robust transmit optimization for multiuser {MISO} downlinks:
  {T}ractable approximations by conic optimization,'' available online at
  {\urlstyle{tt} \url{http://arxiv.org/abs/1108.0982}}.

\bibitem{QLI_ISPACS10}
Q.~Li and W.-K. Ma, ``Optimal transmit design for {MISO} secrecy-rate
  maximization with general covariance constraints,'' in \emph{Intl. Symp.
  Intelligent Signal Process. and Commun. Syst.}, Dec. 2010.

\end{thebibliography}


\end{document}